\documentclass[aps,pra,twocolumn,showpacs,superscriptaddress]{revtex4-1}

\usepackage{graphicx}
\usepackage[colorlinks,citecolor=blue,urlcolor=blue,linkcolor=blue]{hyperref}
\usepackage{siunitx}	%for example \SI{5.0}{\micro\meter}
\usepackage{amsmath}
\usepackage{amssymb}
\usepackage{amsthm}
\usepackage{braket}
\usepackage{booktabs}
\usepackage{footmisc}
\usepackage{mathrsfs}
\usepackage{cleveref}
\newcommand{\op}[1]{\hat{#1}}
\newcommand{\opd}[1]{\hat{#1}^{\dagger}}

\newcommand{\unit}{1\!\!1}

\newcommand{\sgn}{\mathrm{sgn}}
\newcommand{\modu}{\mathrm{mod}}
\newcommand{\perm}{\mathrm{perm}}
\newcommand{\abs}[1]{\left| #1 \right|} % for absolute value
\newcommand{\bracket}[2]{\ensuremath{\left\langle#1 \vphantom{#2}\right| \left. #2\vphantom{#1}\right\rangle}}

\theoremstyle{plain}

\theoremstyle{definition}

\begin{document}
\title{Totally Destructive Interference for Permutation-Symmetric Many-Particle States}

\author{Christoph Dittel}
\email{christoph.dittel@uibk.ac.at}
\affiliation{Institut f{\"u}r Experimentalphysik, Universit{\"a}t Innsbruck, Technikerstr. 25, 6020 Innsbruck, Austria}
\affiliation{Physikalisches Institut, Albert-Ludwigs-Universit{\"a}t Freiburg, Hermann-Herder-Str. 3, 79104 Freiburg, Germany}

\author{Gabriel Dufour}
%\email{gabriel.dufour@physik.uni-freiburg.de}
\affiliation{Physikalisches Institut, Albert-Ludwigs-Universit{\"a}t Freiburg, Hermann-Herder-Str. 3, 79104 Freiburg, Germany}
\affiliation{Freiburg Institute for Advanced Studies, Albert-Ludwigs-Universit{\"a}t-Freiburg, Albertstr. 19, 79104 Freiburg, Germany}

\author{Mattia Walschaers}
\affiliation{Laboratoire Kastler Brossel, UPMC-Sorbonne Universit\'{e}s, ENS-PSL Research University, Coll\`{e}ge de France, CNRS; 4 place Jussieu, 75252 Paris, France}

\author{Gregor Weihs}
%\email{gregor.weihs@uibk.ac.at}
\affiliation{Institut f{\"u}r Experimentalphysik, Universit{\"a}t Innsbruck, Technikerstr. 25, 6020 Innsbruck, Austria}

\author{Andreas Buchleitner}
%\email{abu@uni-freiburg.de}
\affiliation{Physikalisches Institut, Albert-Ludwigs-Universit{\"a}t Freiburg, Hermann-Herder-Str. 3, 79104 Freiburg, Germany}

\author{Robert Keil}
%\email{robert.keil@uibk.ac.at}
\affiliation{Institut f{\"u}r Experimentalphysik, Universit{\"a}t Innsbruck, Technikerstr. 25, 6020 Innsbruck, Austria}

%----------------------------------------------------------------------------------
%                            Abstract
%----------------------------------------------------------------------------------
\date{\today}

\begin{abstract}
Several distinct classes of unitary mode transformations have been known to exhibit the strict suppression of a large set of transmission events, as a consequence of totally destructive many-particle interference. In another work [Dittel et al., Phys. Rev. Lett. \textbf{120}, 240404 (2018)] we unite these cases by identifying a general class of unitary matrices which exhibit such interferences. Here, we provide a detailed theoretical analysis that substantially expands on all aspects of this generalisation: We prove the suppression laws put forward in our other paper, establish how they interrelate with forbidden single-particle transitions, show how all suppression laws hitherto known can be retrieved from our general formalism, and discuss striking differences between bosons and fermions. Furthermore, beyond many-particle Fock states on input, we consider arbitrary pure initial states and derive suppression laws which stem from the wave function's permutation symmetry alone. Finally, we identify conditions for totally destructive interference to persist when the involved particles become partially distinguishable. 
\end{abstract}


\maketitle%----------------------------------------------------------------------------------
%            Introduction
%----------------------------------------------------------------------------------
\section{Introduction}
Based on the permutation symmetry of their many-particle wave function, the symmetrization postulate \cite{Heisenberg-MR-1926,Dirac-TQ-1926,Messiah-SP-1964,Girardeau-PS-1965} fundamentally distinguishes between two types of quanta: bosons and fermions. Its consequences are profound for many areas of physics. In the context of many-particle interference, the wave function's symmetry  impacts the dynamics such as to manifest distinct signatures in the counting statistics. The most prominent scenario involves two identical, non-interacting particles, initially occupying different modes of a balanced two-mode coupler. In the case of indistinguishable bosons \cite{Hong-MS-1987,Shih-NT-1988,Kaufman-TW-2014,Lopes-AH-2015}, the probability amplitudes for both bosons being transmitted and both being reflected cancel each other perfectly. This totally destructive two-particle interference forces both particles to end up in the same final mode. In contrast, for indistinguishable fermions \cite{Liu-QI-1998,Kiesel-OH-2002}, the interference causes both particles to occupy different final modes, in accordance with Pauli's exclusion principle \cite{Pauli-ZA-1925,Pauli-EP-1964}. Similar behaviour has been observed for an increasing number of bosons \cite{Ou-OF-1999,Xiang-DT-2006,Niu-OG-2009,Ra-NQ-2013,Dufour-MP-2017}, and the emergence of totally destructive interference of bosons has also been investigated in three \cite{Mattle-NC-1995,Weihs-TP-1996,Campos-TP-2000,Meany-NC-2012,Spagnolo-TP-2013} and four \cite{Mattle-NC-1995,Meany-NC-2012} mode setups as well as in the absence of scattering elements \cite{Maehrlein-HOM-2017}. For many-particle states and an arbitrary number of modes, several specific scattering scenarios giving rise to totally destructive many-particle interference have been identified: the discrete Fourier transformation \cite{Lim-GH-2005,Tichy-ZT-2010,Tichy-MP-2012,Carolan-UL-2015,Crespi-SL-2016,Su-MI-2017}, the $J_x$ unitary \cite{Perez-Leija-CQ-2013,Weimann-IQ-2016}, Sylvester interferometers \cite{Crespi-SL-2015,Viggianiello-EG-2017} and hypercube unitaries \cite{Dittel-MB-2017}. In all these cases, the underlying unitary transformation possesses a highly symmetric structure. Then, for initial particle configurations which satisfy a closely related symmetry condition, specific output configurations occur with zero probability, due to a perfect cancellation of all contributing many-particle transition amplitudes. This has led to the formulation of so called \textit{suppression laws} \cite{Lim-GH-2005,Tichy-ZT-2010,Tichy-MP-2012,Crespi-SL-2015,Dittel-MB-2017,Viggianiello-EG-2017}, which provide a sufficient condition for the identification of  forbidden output events with little computational expense. The symmetry properties of the unitary are thus exploited to circumvent the computationally expensive addition of all contributing many-particle amplitudes \cite{Mayer-CS-2011}. 

So far, it has remained an open question whether all those suppression laws could be understood as the consequence of one common symmetry property, and inferred from a general condition for totally destructive interference. Furthermore, as a consequence of the wave function's symmetry, suppression laws generally differ for bosons and fermions. Nevertheless, it was noticed in \cite{Tichy-MP-2012} and \cite{Dittel-MB-2017} that the fermionic suppression law can, under certain circumstances, induce the same suppressed output events as in the bosonic case. No reason for this behaviour could hitherto be identified.  Finally, most previous studies on many-particle interference considered product input states, however, two-particle entangled states can mimic bosonic as well as fermionic interference on the balanced two-mode coupler \cite{Michler-IB-1996,Mattle-DC-1996,Zeilinger-QE-1998,Tichy-EI-2013}, depending on their internal phase. This can be generalised for arbitrary particle and mode numbers \cite{Matthews-OF-2013} and suggests that totally destructive interference must also occur for many-particle input states that cannot be expressed as product states. Whether suppression laws exist for such states that are not necessarily totally (anti-)symmetric under permutation, is unknown to date.

In Ref.~\cite{Dittel-TDletter-2017} we formulate a criterion which encompasses all known suppression laws and, for any given initial product state, pinpoints unitary transformation matrices which display totally destructive many-particle interference. These transformation matrices are largely determined by the eigenbases of permutations which leave the initial particle configuration invariant. The suppression of many-particle output events does not depend on a particular choice of eigenbasis, and the suppression criterion can be recast in terms of the eigenvalues of the permutation under consideration.

In Sec.~\ref{sec:supplaw} of the present work, we provide a comprehensive theoretical framework for totally destructive interference, which extends the concepts presented in \cite{Dittel-TDletter-2017} and allows to prove the general bosonic suppression law. For fermions, we present two different suppression laws: The first is an adaptation of the bosonic suppression law, which relies only on the wave function's permutation symmetry and
provides a sufficient condition for suppression, closely resembling the bosonic case. The second provides a more comprehensive sufficient condition for ouput event suppression if one additionally assumes product input states. Moreover, we also elaborate on single-particle dynamics, which likewise can lead to forbidden output events even in the absence of many-particle interference. While we first derive our suppression laws in the traditional scattering matrix approach \cite{Mayer-CS-2011}, we further show that they can be consistently inferred from the input state's permutation symmetry.

The applicability of our suppression laws to all hitherto known cases is demonstrated in Sec.~\ref{sec:Applications}. For the discrete Fourier transformation \cite{Tichy-MP-2012} the bosonic suppression law is recovered, while our generalised suppression law goes beyond the hitherto known fermionic result. We further retrieve the suppression laws for Sylvester interferometers \cite{Crespi-SL-2015,Viggianiello-EG-2017} and hypercube unitaries \cite{Dittel-MB-2017}. For the $J_x$ unitary, the two-boson suppression law \cite{Weimann-IQ-2016} is generalised to arbitrary (bosonic or fermionic) particle numbers.

In Sec.~\ref{sec:purestates}, we then extend our analysis to arbitrary pure input states. We only require the initial state to be permutation-symmetric, to derive a suppression law which determines unitaries and associated forbidden output events. The versatility of this approach is highlighted by examples that treat superpositions of indistinguishable particles, as well as entangled many-particle states. Finally, we address partially distinguishable particles, investigate under which conditions the suppression laws for indistinguishable particles remain unaffected, and confirm the \textit{zero-probability conjecture} formulated in \cite{Shchesnovich-PI-2015}.

%----------------------------------------------------------------------------------
%            Indistinguishable Particles in Fock-States
%----------------------------------------------------------------------------------
\section{Suppression laws for Fock product states}
\label{sec:supplaw}
%----------------------------------------------------------------------------------
%            Preliminaries
%----------------------------------------------------------------------------------
\subsection{Preliminaries}
Consider the coherent evolution of $N$ identical and non-interacting particles that can be distributed among $n$ modes. The creation operator associated with a particle in the $j$th input (output) mode is denoted by $\opd{a}_j$ ($\opd{b}_j$), where $j\in\{1,\dots,n\}$. Our notation does not discern between bosonic and fermionic creation operators in the following. However, note that the usual (anti-) commutation relations apply, that is
\begin{align}\label{eq:com.bosons}
[\op{a}_j,\opd{a}_k]=\delta_{j,k}\ \ ,\ \ [\op{a}_j,\op{a}_k]=[\opd{a}_j,\opd{a}_k]=0
\end{align}
for bosonic, and
\begin{align}\label{eq:com.fermions}
\{\op{a}_j,\opd{a}_k\}=\delta_{j,k}\ \ ,\ \ \{\op{a}_j,\op{a}_k\}=\{\opd{a}_j,\opd{a}_k\}=0
\end{align}
for fermionic operators, and likewise for $\opd{b}_j$ and $\op{b}_j$. 

We denote initial many-particle configurations with $r_j$ particles in the $j$th mode either by their mode occupation list $\vec{r}=(r_1,\dots,r_n)$ or by their mode assignment list $\vec{d}(\vec{r})=(d_1(\vec{r}),\dots,d_N(\vec{r}))$, with $d_\alpha(\vec{r})$ specifying the mode number of the $\alpha$th particle, with $\alpha\in\{1,\dots,N\}$. For indistinguishable particles, the ordering in $\vec{d}(\vec{r})$ is irrelevant and, unless otherwise stated, $\vec{d}(\vec{r})$ is given in ascending order of the modes. The initial bosonic (B)/fermionic (F) \textit{Fock product state} defined by $\vec{r}$ reads \cite{Tichy-II-2014}
\begin{align}\label{eq:state}
\begin{split}
\ket{\Psi^\mathrm{B/F}(\vec{r})}&=\prod_{j=1}^n\frac{(\opd{a}_j)^{r_j}}{\sqrt{r_j!}}\ket{0}\\
&=\frac{1}{\sqrt{\prod_{j=1}^n r_j!}}\prod_{\alpha=1}^N\opd{a}_{d_\alpha(\vec{r})}\ket{0},
\end{split}
\end{align}
with $\ket{0}$ the vacuum state. Analogously, final particle configurations are either denoted by occupation or assignment lists $\vec{s}$ and $\vec{d}(\vec{s})$, respectively, and the corresponding final state reads
\begin{align}\label{eq:statefinal}
\ket{\Psi^\mathrm{B/F}(\vec{s})}&=\prod_{j=1}^n\frac{(\opd{b}_j)^{s_j}}{\sqrt{s_j!}}\ket{0}.
\end{align}
The fermionic anticommutaion relation~\eqref{eq:com.fermions} in Eq.~\eqref{eq:state} immediately implies that multiple mode occupation is forbidden for indistinguishable fermions, in accordance with Pauli's exclusion principle \cite{Pauli-ZA-1925}. 

The evolution from initial to final states is modelled by a unitary matrix $U$. In the Heisenberg picture, creation operators transform according to \cite{Tichy-MP-2012}
\begin{align}\label{eq:atransform}
\opd{a}_j\rightarrow \sum_{k=1}^n U_{j,k}\opd{b}_k\ ,
\end{align}
leading to the mapping $\ket{\Psi^\mathrm{B/F}(\vec{r})}\mapsto  \ket{\Psi_\mathrm{evo}^\mathrm{B/F}(\vec{r})}$. In~\eqref{eq:atransform}, $U_{j,k}$ specifies the single particle transition amplitude from input mode $j$ to output mode $k$. Born's rule then gives the probability to detect the final particle configuration $\vec{s}$ as
\begin{align}\label{eq:transprob}
P_\mathrm{B/F}(\vec{r},\vec{s},U)=\abs{\bracket{\Psi^\mathrm{B/F}(\vec{s})}{\Psi_{\mathrm{evo}}^\mathrm{B/F}(\vec{r})}}^2.
\end{align}

It is common \cite{Lim-GH-2005,Tichy-ZT-2010,Tichy-MP-2012,Tichy-II-2014,Crespi-SL-2015,Dittel-MB-2017} to express the transition probability~\eqref{eq:transprob} for bosons and fermions in terms of the permanent $\perm(M)$ and determinant $\det(M)$ of the \textit{scattering matrix} $M$, respectively \cite{Mayer-CS-2011}. The elements of this matrix are determined by the initial and final particle configurations, according to $M_{\alpha,\beta}=U_{d_\alpha(\vec{r}),d_\beta(\vec{s})}$, for all $\alpha, \beta\in\{1,...,N\}$. By evaluation of the scalar product in Eq.~\eqref{eq:transprob} one  finds
\begin{align}\label{eq:perm}
P_\mathrm{B}(\vec{r},\vec{s},U)=\frac{1}{\prod_{j=1}^n r_j!s_j!}\abs{\perm(M)}^2
\end{align}
for bosons, and
\begin{align}\label{eq:det}
P_\mathrm{F}(\vec{r},\vec{s},U)=\abs{\det(M)}^2
\end{align}
for fermions \cite{Mayer-CS-2011}.  On the other hand, in the case of distinguishable particles (D), one has to sum over all many-particle transition probabilities, which results in
\begin{align}\label{eq:distperm}
P_\mathrm{D}(\vec{r},\vec{s},U)=\frac{1}{\prod_{j=1}^n s_j!}\perm(\abs{M}^2)
\end{align}
with $\abs{.}^2$ the squared modulus of individual matrix elements.

%----------------------------------------------------------------------------------
%            Unitary Transformation Matrices
%----------------------------------------------------------------------------------
\subsection{Unitary transformation matrices}\label{sec:Unitarymatrices}
Following our approach in \cite{Dittel-TDletter-2017}, we now identify those unitary matrices $U$ that exhibit totally destructive many-particle interferences, given a Fock input state~\eqref{eq:state} defined by its mode occupation list $\vec{r}$. We start out from any permutation operation $\mathscr{P}$ that leaves $\vec{r}$ invariant,
\begin{align}\label{eq:rinvariant}
\mathscr{P}\vec{r}=\vec{r},
\end{align}
with $\mathscr{P}_{j,k}=\delta_{\pi(j),k}$, $\pi \in \mathrm{S}_n$, and $\mathrm{S}_n$  the symmetric group on the set of modes $\{1,\dots,n\}$. An eigendecomposition $\mathscr{P}=ADA^\dagger$ then extracts the eigenvectors of $\mathscr{P}$ as the columns of the unitary matrix $A\in\mathbb{C}^{n\times n}$, and the associated eigenvalues $\lambda_j$ as entries of the diagonal matrix $D=\mathrm{diag}(\lambda_1,\dots,\lambda_n)$. Since $\mathscr{P}$ is unitary, the $\lambda_j$ live on the unit circle and appear as phase-factors further down. 

We now construct evolution matrices of the form
\begin{align}\label{eq:UA}
U=\Theta~A~\Sigma,
\end{align} 
with arbitrary, diagonal unitary matrices $\Theta\in\mathbb{C}^{n\times n}$ and $\Sigma\in\mathbb{C}^{n\times n}$, which describe local phase operations on the input and output modes, respectively. For Fock states of the form~\eqref{eq:state}, many-particle interference is insensitive to these local phases, and the only relevant transformation is induced by the matrix $A$, composed of the eigenstates of $\mathscr{P}$. 

The fundamental reason for choosing $U$ according to Eq.~\eqref{eq:UA} is anchored in its permutation characteristics. In particular, $U$ is invariant under $\mathscr{P}$ up to a multiplication of each column with the respective eigenvalue contained in $D$, and the imprinting of local phases encoded in the diagonal unitary matrix $Z=\mathscr{P}~\Theta~\mathscr{P}^\dagger~\Theta^\dagger$, such that
\begin{align}
\mathscr{P}~U=Z~U~D. \label{eq:MatrixSym}
\end{align}
The unitarity of $Z$ is due to $\mathscr{P}$ and $\Theta$ being unitary, and its diagonality results from $\mathscr{P}$ being a permutation operator. In element-wise notation, Eq.~\eqref{eq:MatrixSym} reads, for all $j,k\in\{1,\dots,n\}$
\begin{align}\label{eq:sym}
U_{\pi(j),k}=\exp\left(i[\theta(\pi(j))-\theta(j)]\right)U_{j,k}\lambda_k 
\end{align}
with $\Theta_{j,j}=e^{i\theta(j)}$. Since $\pi$ is bijective, \eqref{eq:sym} establishes a \textit{symmetric phase relation} between the matrix elements $U_{j,k}$ and those, $U_{\pi(j),k}$, of its image under $\mathscr{P}$. Since local phases $\theta$ cannot affect many-particle interference, the latter must be controlled by the eigenvalues $\lambda_k$ of $\mathscr{P}$. In particular, for a given output event $\vec{s}$, all the relevant eigenvalues are collected in the \textit{final eigenvalue distribution} $\Lambda(\vec{s})=\{\lambda_{d_1(\vec{s})},\dots, \lambda_{d_N(\vec{s})}\}$, given that the transition probabilities~\eqref{eq:perm} and~\eqref{eq:det} are governed by the elements $U_{d_\alpha(\vec{r}),d_\beta(\vec{s})}$. It contains all $N$ eigenvalues $\Lambda_\beta(\vec{s})=\lambda_{d_\beta(\vec{s})}$ associated with the final mode assignment list $\vec{d}(\vec{s})$ and is a multiset, that is, the ordering of elements in $\Lambda(\vec{s})$ is unspecified and equal eigenvalues can occur multiple times.

Further characteristics of the unitaries \eqref{eq:UA}, associated with $\vec{r}$, follow from the cycle decomposition \cite{Rotman-TG-1973} of the underlying permutation:
Let $\pi\in\mathrm{S}_n$ consist of cycles with $L$ different lengths $m_1,\dots,m_L$, such that the period of the permutation is given by the least common multiple ($\mathrm{lcm}$) of all cycle lengths: $\mathscr{P}^m=\unit$ for $m=\mathrm{lcm}(m_1,\dots,m_L)$ \cite{Rotman-TG-1973}. Accordingly, each cycle with length $m_l$ contributes the set $\{e^{i\frac{2\pi}{m_l}},e^{i\frac{2\pi}{m_l}2},\dots,e^{i\frac{2\pi}{m_l}m_l}\}$ to the eigenvalues of $\mathscr{P}$. 

For example, consider the permutation 
\begin{align*}
\pi&=\begin{pmatrix}1&2&3&4&5&6&7&8&9&10&11\\ 2&3&1&5&6&4&8&9&7&11&10\end{pmatrix}\\
&=(1~2~3)(4~5~6)(7~8~9)(10~11)
\end{align*}
with the cycle decomposition of $\pi$ in the second line. This permutation has $L=2$ different cycle lengths $m_1=3$ and $m_2=2$, and is of order $m=\mathrm{lcm}(2,3)=6$. The eigenvalues of the corresponding permutation operator read $\{e^{i\frac{2\pi}{3}},e^{i\frac{4\pi}{3}},1,e^{i\frac{2\pi}{3}},e^{i\frac{4\pi}{3}},1,e^{i\frac{2\pi}{3}},e^{i\frac{4\pi}{3}},1,-1,1\}$ since $\pi$ consists of $3$ cycles of length $m_1=3$, and one cycle of length $m_2=2$. 

The ``canonical'' matrix $A_\mathrm{C}$ which diagonalizes $\mathscr{P}$ is of block-diagonal form, up to a permutation of the rows (tantamount to relabelling the input modes). Each block corresponds to a cycle of $\pi$ and, with $m_l$ the cycle length, consist of a $m_l\times m_l$ Fourier matrix (defined in Eq.~\eqref{eq:UFT} below). Modulo permutations of the columns (tantamount to relabelling the output modes), any eigenbasis $A$ of $\mathscr{P}$ can then be obtained from $A_\mathrm{C}$ by rotations in the degenerate subspaces. This partially washes out the block structure as the rotations mix all columns (eigenvectors of $\mathscr{P}$) with equal eigenvalue. However, the $j$th component $A_{j,k}$ of the $k$th eigenvector with eigenvalue $\lambda_k$ and, by \eqref{eq:UA}, the matrix element $U_{j,k}$, are necessarily zero if all eigenvectors of $A_\mathrm{C}$ with eigenvalue $\lambda_k$ have a vanishing $j$th component. This is the case if $j$ is in a cycle of length $m_l$ and $\lambda_k^{m_l} \neq 1$. The symmetric phase relation~\eqref{eq:sym} encodes this characteristic of $U$ since for a mode $j$ which belongs to a cycle of $\pi$ with length $m_l$, one obtains
\begin{align}\label{eq:mlassigned}
U_{j,k}=U_{\pi^{m_l}(j),k}=U_{j,k}\ \lambda_k^{m_l},
\end{align}
such that $U_{j,k}=0$ unless $\lambda_k^{m_l}=1$. These zero-entries in the single-particle transition matrix inevitably lead to forbidden particle transition events on the level of single-particle dynamics, as we show in the following.

%----------------------------------------------------------------------------------
%            Zero transmission due to single-particle dynamics
%----------------------------------------------------------------------------------
\subsection{Forbidden events due to single-particle dynamics}
\label{sec:zerotransmission}

\begin{figure}[t]
\centering
\includegraphics[width=0.38\textwidth]{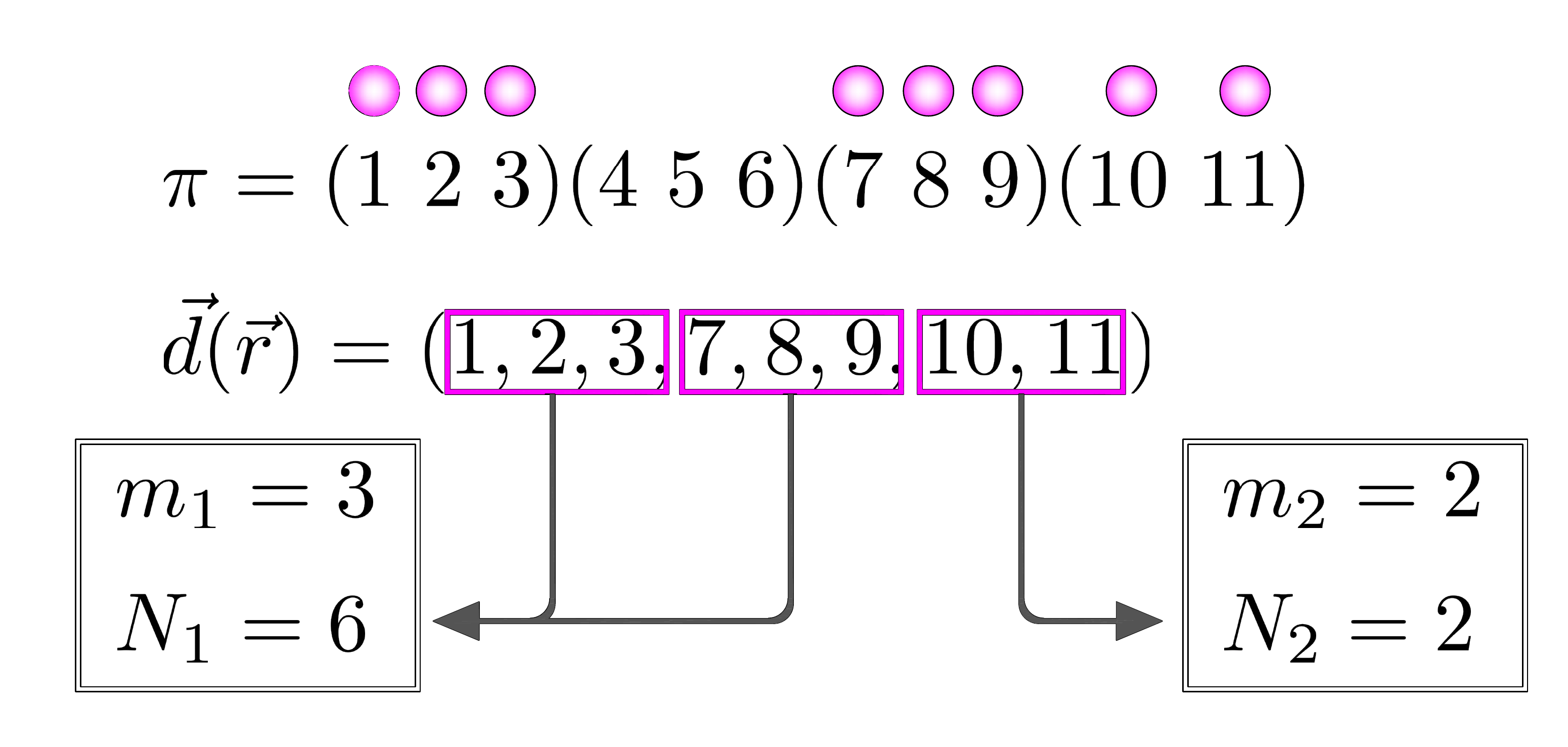} 
\caption{(Color online) Example of a cycle decomposition. Magenta balls indicate the initial occupation in cycles of $\pi$ according to the particle configuration defined by the input state's mode assignment list $\vec{d}(\vec{r})$. Black arrows indicate how the initially occupied cycles contribute to the total particle numbers $N_1$ and $N_2$ in cycles with lengths $m_1$ and $m_2$, respectively.}
\label{fig:singleparticledynamics}
\end{figure}

In order to identify those transmission events which are forbidden as a consequence of the underlying single-particle dynamics, we consider the initial population of cycles of $\pi$: Given that $\vec{r}$ is invariant under $\pi$, the number of particles in each mode of a cycle must be equal, so that the total number of particles in all modes of a single cycle with length $m_l$ is an integer multiple of $m_l$. Since, for given $\vec{r}$ and $\pi$, there can be many cycles with the same length, we denote the total number of particles that are initially prepared in cycles with length $m_l$ by $N_l$, where $l\in\{1,\dots,L\}$, such that $\sum_{l=1}^LN_l=N$. This is illustrated in Fig.~\ref{fig:singleparticledynamics}, for the permutation considered above, and for the initial particle configuration $\vec{d}(\vec{r})=(1,2,3,7,8,9,10,11)$, with $N_1=6$ and $N_2=2$ particles in modes associated with cycles of length $m_1=3$ and $m_2=2$, respectively.

By virtue of Eq.~\eqref{eq:mlassigned}, a particle prepared in mode $j$, associated with a cycle of length $m_l$, will always end up in a mode $k$ for which $\lambda_k^{m_l}=1$. Hence, for $N_l$ particles associated with cycles of length $m_l$, taking the $m_l$th power of each element of the final eigenvalue distribution $\Lambda(\vec{s})$ must produce a set with at least $N_l$ elements equal to unity, i.e.
\begin{align}\label{eq:Ndef}
\mathscr{N}_l(\vec{s})=\abs{\{\lambda\in\Lambda(\vec{s})\ :\ \lambda^{m_l}=1\}}\geq N_l,
\end{align}
where $\abs{.}$ denotes the cardinality of a set. Therefore, we find $P(\vec{r},\vec{s},U)=0$ for the transition probability (regardless of the particles' types and mutual distinguishability), whenever, for any $l\in\{1,\dots,L\}$,
\begin{align}\label{eq:suppsingle}
\mathscr{N}_l(\vec{s})<N_l.
\end{align}
A vanishing transition probability can also be characterised as follows: If condition~\eqref{eq:suppsingle} is satisfied, then for every permutation $\sigma\in \mathrm{S}_N$ there exists at least one particle $\alpha$  (which depends on $\sigma$) such that $\alpha$ is initially in a cycle of length $m_l$, but $\lambda_{\sigma(\alpha)}^{m_l}\neq1$. Therefore, all elements $M_{\alpha,\sigma(\alpha)}$ of the scattering matrix vanish, and $\perm(M)=\det(M)=\perm(|M|^2)=0$ (recall Eqs.~(\ref{eq:perm}-\ref{eq:distperm})).

Note that the thus defined forbidden many-particle transitions are a direct manifestation of the vanishing entries of the single-particle transformation matrix $U$, and therefore \emph{independent} of particle type and distinguishability.

%----------------------------------------------------------------------------------
%            Scattering Matrix Approach
%----------------------------------------------------------------------------------
\subsection{The scattering matrix approach}
\label{sec:scatteringMatrix}

We now turn to genuine many-body dynamics as formalized by the scattering matrix expressions in Eq.~\eqref{eq:perm} and \eqref{eq:det}. The permutation characteristics~\eqref{eq:MatrixSym} of the transformation matrices $U$ are directly transferred to the scattering matrix $M$, for which
\begin{align}\label{eq:Msym}
\bar{\mathscr{P}}M=\bar{Z}~M~\bar{D}
\end{align}
with $\bar{\mathscr{P}}_{\alpha,\beta}=\mathscr{P}_{d_\alpha(\vec{r}),d_\beta(\vec{r})}$, $\bar{Z}_{\alpha,\beta}=Z_{d_\alpha(\vec{r}),d_\beta(\vec{r})}$ and the diagonal matrix $\bar{D}=\mathrm{diag}(\lambda_{d_1(\vec{s})},\dots,\lambda_{d_N(\vec{s})})$, with the eigenvalues of the final eigenvalue distribution $\Lambda(\vec{s})$ on its diagonal. Note that, for multiply occupied initial modes, $\bar{\mathscr{P}}$ is \emph{not} a permutation operator and $\bar{Z}$ is \emph{not} diagonal.

\subsubsection{Bosons}\label{sec:bosonsScatter}
First, let us consider indistinguishable bosons and investigate the permanent in Eq.~\eqref{eq:perm}: Given the fact that a permanent is invariant under permutations of its argument's rows and columns \cite{Minc-P-1984}, we find
\begin{align}\label{eq:pmetam}
\perm(\bar{\mathscr{P}}M)=\eta\ \perm(M),
\end{align}
with the constant factor $\eta \neq 0$ accounting for multiply occupied initial modes. On the other hand, $\prod_{\alpha=1}^N \exp(i[\theta(\pi(d_\alpha(\vec{r})))-\theta(d_\alpha(\vec{r})) ])=1$, since the number of particles is equal in all modes corresponding to the same cycle. Therewith, we obtain for the permanent of the matrix product on the right hand side of Eq.~\eqref{eq:Msym}: 
\begin{align*}
\perm(\bar{Z}~M~\bar{D})=\eta\ \perm(M)\prod_{\alpha=1}^N\lambda_{d_\alpha(\vec{s})},
\end{align*}
and, with \eqref{eq:pmetam}, 
\begin{align*}
\eta\ \perm(M)=\eta\ \perm(M)\prod_{\alpha=1}^N\lambda_{d_\alpha(\vec{s})}.
\end{align*}
Hence, only if
\begin{align*}
\prod_{\alpha=1}^N\lambda_{d_\alpha(\vec{s})} =\prod_{\alpha=1}^N\Lambda_\alpha(\vec{s}) = 1
\end{align*}
can $\perm(M)\neq 0$, and, by Eq.~\eqref{eq:perm}, we therefore conclude that $P_\mathrm{B}(\vec{r},\vec{s},U)=0$ if:
\begin{align}\label{eq:bosonsupplaw}
\prod_{\alpha=1}^N\Lambda_\alpha(\vec{s}) \neq 1.
\end{align}
This is a sufficient condition for the suppression of the output event $\vec{s}$, as a consequence of the perfect cancellation of many-particle amplitudes by destructive many-particle interference. By~\eqref{eq:bosonsupplaw}, this interference is unambiguously related to a simple property of the final eigenvalue distribution $\Lambda(\vec{s})=\{\lambda_{d_1(\vec{s})},\dots,\lambda_{d_N(\vec{s})}\}$. 

\begin{figure}[t]
\centering
\includegraphics[width=0.47\textwidth]{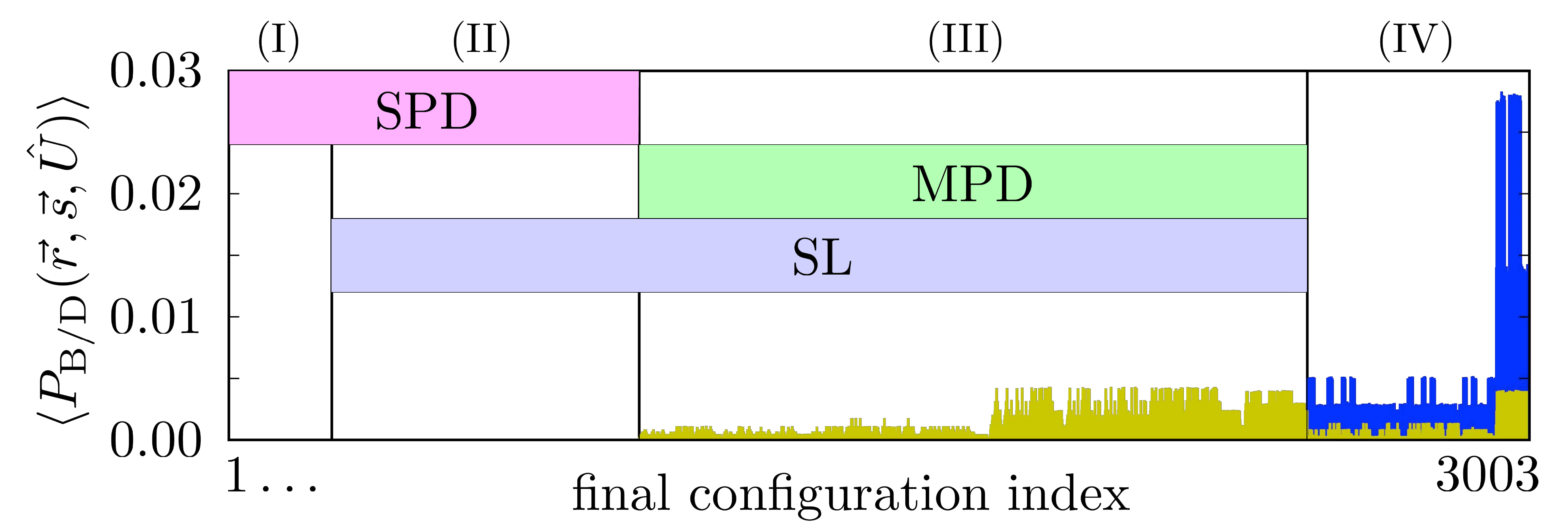} 
\caption{(Color online) Single-particle dynamics and many-particle interference for bosons: For the input state $\vec{d}(\vec{r})=(1,2,3,10,11)$ and $10\,000$ randomly generated eigenbases of the permutation operator associated with $\pi=(1~2~3)(4~5~6)(7~8~9)(10~11)$, the mean transition probability $\langle P_{\mathrm{B/D}}(\vec{r},\vec{s},U)\rangle$ for indistinguishable bosons (B, blue bars) and distinguishable particles (D, yellow bars) is shown for all $3003$ possible output events of $N=5$ particles transmitted across $n=11$ modes. Vanishing transmission probabilities in domains (I) and (II) occur due to single-particle dynamics (SPD). The suppression due to multi-particle dynamics (MPD) in domain (III), as well as of the SPD-suppressed configurations in domain (II), is predicted by the bosonic suppression law (SL). Only output events listed in domain (IV) occur with finite probability for indistinguishable bosons.}
\label{fig:bosons}
\end{figure}

Let us illustrate this relation by an example: We consider an initial particle configuration of $N=5$ bosons, $\vec{d}(\vec{r})=(1,2,3,10,11)$, injected into $n=11$ modes, and the aforementioned permutation $\pi=(1~2~3)(4~5~6)(7~8~9)(10~11)$ which leaves $\vec{r}$ invariant. We numerically generate $10\,000$ eigenbases of the associated $\mathscr{P}$, by random rotations in the degenerate subspaces. According to our discussion at the end of Sec.~\ref{sec:Unitarymatrices} above, each $q$-fold degenerate eigenvalue allows us to rotate $A$ by a $q\times q$ unitary matrix, thus mixing the columns of $A$ which are (and remain) associated with this eigenvalue. We calculate the mean transition probability $\langle P_{\mathrm{B/D}}(\vec{r},\vec{s},U)\rangle$, averaged over all realisations of the so-constructed random unitary, for both indistinguishable bosons and distinguishable particles. The results are shown in Fig.~\ref{fig:bosons}, and can be grouped into four different domains: Output events listed in domains (I) and (II) obey condition~\eqref{eq:suppsingle} and are forbidden by single particle dynamics, for indistinguishable bosons, as well as for distinguishable particles. For indistinguishable bosons, all particle configurations listed in domain (III) are suppressed due to totally destructive many-particle interference, while they occur with non-zero probability for distinguishable particles. The suppression law~\eqref{eq:bosonsupplaw} predicts \emph{all} suppressed events in domains (III) and (II). Domain (IV) collects all transmission events which satisfy neither condition~\eqref{eq:suppsingle} nor the suppression law~\eqref{eq:bosonsupplaw}. In our present example, all these configurations occur with non-vanishing probability, for distinguishable as well as for indistinguishable particles.

\subsubsection{Fermions}

Next, we consider indistinguishable fermions and, as imposed by Pauli's principle, only those particle configurations with at most singly occupied modes.  Calculating the determinant of the left hand side of Eq.~\eqref{eq:Msym} reveals
\begin{align}\label{eq:detMsgn}
\det(\bar{\mathscr{P}}M)=(-1)^w \det(M),
\end{align}
where $w$ denotes the number of transpositions (exchanges of two particles) required to permute $\vec{r}$ according to $\mathscr{P}$. Note that $\bar{\mathscr{P}}$ is a permutation operator, since we only consider singly occupied modes in the present, fermionic, case. Analogously to the case of bosons, we obtain for the right hand side of Eq.~\eqref{eq:Msym}
\begin{align}\label{eq:detZMD}
\det(\bar{Z}~M~\bar{D})=\det(M)\prod_{\alpha=1}^N\lambda_{d_\alpha(\vec{s})}.
\end{align}
Equations~\eqref{eq:detMsgn},~\eqref{eq:detZMD} and \eqref{eq:Msym} immediately imply that $\det(M)$ and thus $P_\mathrm{F}(\vec{r},\vec{s},U)$ must vanish if 
\begin{align}\label{eq:Fadaption}
\prod_{\alpha=1}^N\Lambda_\alpha(\vec{s}) \neq (-1)^w.
\end{align}
In the following, we refer to this condition as the \textit{adapted suppression law} for fermions. This terminology will become clear in Sec.~\ref{sec:Bosons} below.

There are, however, more forbidden transmission events predicted neither by condition~\eqref{eq:Fadaption} nor by single-particle dynamics according to Eq.~\eqref{eq:suppsingle}. We now derive a condition which also accounts for those: Given the decomposition ~\eqref{eq:UA} of the underlying unitary, we can write the scattering matrix as
\begin{align}\label{eq:MTAS}
M=\bar{\Theta}~\bar{A}~\bar{\Sigma},
\end{align}
with the matrix elements $\bar{\Theta}_{\alpha,\beta}=\Theta_{d_\alpha(\vec{r}),d_\beta(\vec{r})}$, $\bar{A}_{\alpha,\beta}=A_{d_\alpha(\vec{r}),d_\beta(\vec{s})}$ and $\bar{\Sigma}_{\alpha,\beta}=\Sigma_{d_\alpha(\vec{s}),d_\beta(\vec{s})}$. Note that $\bar{A}$ is composed of eigenvectors of $\bar{\mathscr{P}}$, but does not necessarily form an eigenbasis. Taking the determinant on both sides of Eq.~\eqref{eq:MTAS} then yields 
\begin{align*}
\det(M)&=\det(\bar{\Theta}) \det(\bar{A}) \det(\bar{\Sigma}) \\
&\propto \det(\bar{A}),
\end{align*}
where $\det(\bar{\Theta})\neq 0$ and $\det(\bar{\Sigma}) \neq 0$, since $\bar{\Theta}$ and $\bar{\Sigma}$ are diagonal matrices and have non-zero entries on the diagonal. Consequently  \cite{Lang-LA-1987},
\begin{align*}
 \det(M)\neq 0  \Leftrightarrow \det(\bar{A})\neq 0 \Leftrightarrow \bar{A}\ \  \text{invertible}.
\end{align*} 
Having in mind that only singly occupied modes are considered, it is straightforward to verify $\bar{\mathscr{P}}\ \bar{A}=\bar{A}\ \bar{D}$.  Therefore, if $\bar{A}$ is invertible, then
\begin{align}\label{eq:ADA-1}
\bar{\mathscr{P}}=\bar{A}\ \bar{D}\ \bar{A}^{-1},
\end{align}
and $\bar{\mathscr{P}}$ and $\bar{D}$ have the same spectrum. In other words, if $\det(M)\neq 0$, then $\bar{\mathscr{P}}$ and $\bar{D}$ have the same spectrum. The spectrum of $\bar{D}$ is given by the final eigenvalue distribution $\Lambda(\vec{s})$. On the other hand, the spectrum of $\bar{\mathscr{P}}$ depends on the initial particle distribution over the cycles of $\pi$  (see Sec.~\ref{sec:zerotransmission}): Each initially occupied cycle with length $m_l$ gives rise to a set of eigenvalues $e^{i2\pi\frac{k}{m_l}}$ with $k=1,\dots,m_l$, and the spectrum of $\bar{\mathscr{P}}$ is given by the multiset sum of these sets  \footnote{For example, the multiset sum of $X=\{e^{i\frac{2\pi}{3}},e^{i\frac{4\pi}{3}},1 \}$ and $Y=\{-1,1\}$ is given by $X\uplus Y=\{e^{i\frac{2\pi}{3}},e^{i\frac{4\pi}{3}},1,-1,1\}$}. For convenience, we denote the spectrum of $\bar{\mathscr{P}}$ as the \textit{initial eigenvalue distribution} $\Lambda_{\mathrm{ini}}$. Thus, by contraposition, if
\begin{align}\label{eq:suppcondF} 
\Lambda(\vec{s})\neq \Lambda_\mathrm{ini},
\end{align}
then $\det(M)= 0$, and, consequently, $P_\mathrm{F}(\vec{r},\vec{s},U)=0$.

By condition~\eqref{eq:suppcondF}, which we will hereafter refer to as the \textit{extended suppression law}, we obtained a more comprehensive suppression law for fermionic initial many-particle product states. As we elaborate in App.~\ref{appsec:proof1}, this condition covers all output events which are suppressed according to condition~\eqref{eq:Fadaption}. We conclude that fermions feature a suppression that is, remarkably,  restricting the set of allowed transmission events more tightly than in the bosonic variant. Mathematically, this behaviour can be traced back to the anticommutativity of fermionic creation operators, which induces the determinant that emerges in the evaluation of the scalar product in Eq.~\eqref{eq:transprob}. While the adapted suppression law~\eqref{eq:Fadaption} conditions the suppression of the output events $\vec{s}$ on the product of the entries of the final eigenvalue distribution, the extended suppression law~\eqref{eq:suppcondF} specifies the eigenvalue distribution of allowed transmission events element-wise, which constitutes a more stringent criterion.

\begin{figure}[t]
\centering
\includegraphics[width=0.47\textwidth]{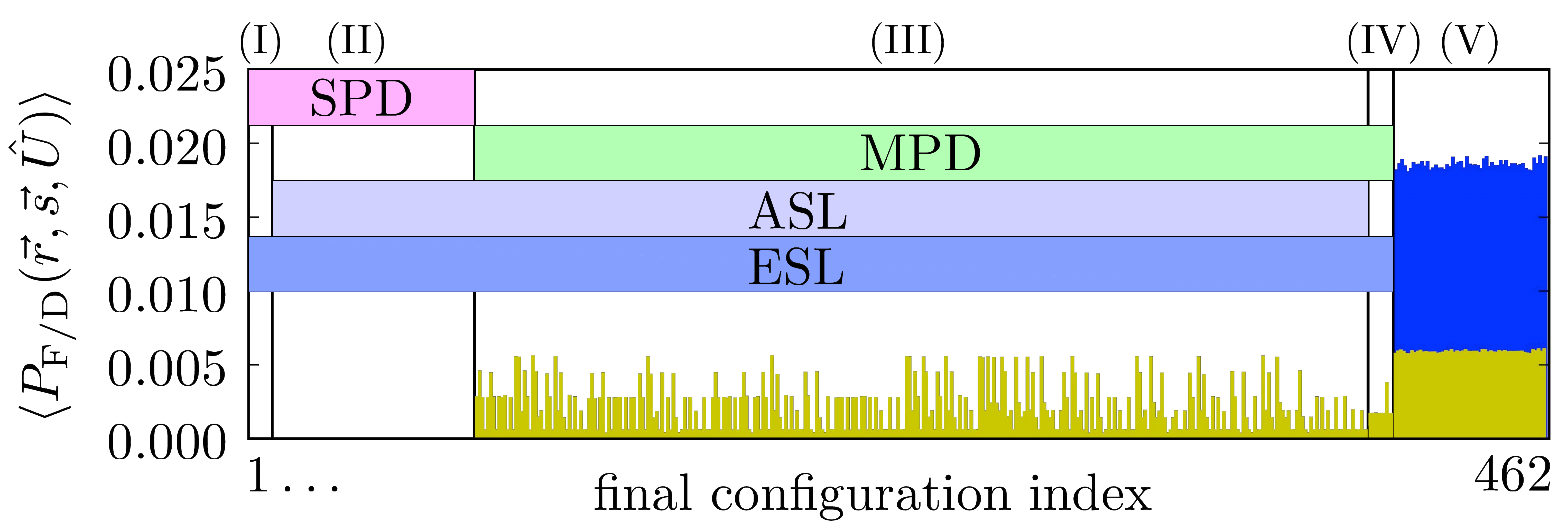} 
\caption{(Color online) Single-particle dynamics and many-particle interference for fermions: For the same setting as in Fig.~\ref{fig:bosons}, the mean transition probability $\langle P_{\mathrm{F/D}}(\vec{r},\vec{s},U)\rangle$ for indistinguishable fermions (F, blue bars) and distinguishable particles (D, yellow bars) is shown, with the latter renormalised to all $462$ possible output events with at most singly occupied modes. Domains (I) and (II) collect all output events which are forbidden on the basis of single-particle dynamics (SPD), while configurations in domain (III) and (IV) are suppressed due to destructively interfering multi-particle dynamics (MPD). The adapted suppression law (ASL) identifies suppressed events in domains (II) and (III), while the extended suppression law (ESL) grasps the suppression of \emph{all} configurations in domains (I)-(IV). Output events in domain (V) are not concerned by the suppression laws, and occur with finite probability.}
\label{fig:fermions}
\end{figure}

Note that the extended suppression law~\eqref{eq:suppcondF} is not simply a combination of the adapted suppression law~\eqref{eq:Fadaption} with that~\eqref{eq:suppsingle} for forbidden transmission events due to single-particle dynamics, as one can appreciate with in the following example: Let us consider the same scenario as discussed for bosons at the end of Sec.~\ref{sec:bosonsScatter}: For the input state $\vec{r}=(1,1,1,0,0,0,0,0,0,1,1)$ we generate $10\,000$ random eigenbases of the permutation operator which represents $\pi=(1~2~3)(4~5~6)(7~8~9)(10~11)$, following the same procedure as above. Figure~\ref{fig:fermions} lists the resulting mean transition probabilities $\langle P_{\mathrm{F/D}}(\vec{r},\vec{s},U)\rangle$ for indistinguishable fermions and for distinguishable particles, where the latter is renormalised to all possible output events with at most singly occupied modes, for better comparability. 

All final particle configurations are grouped into five different domains: Domains (I) and (II) list all configurations which are forbidden due to single-particle dynamics, according to condition~\eqref{eq:suppsingle}. Suppressed output events due to many-particle interference are grouped in domains (III) and (IV). While the adapted suppression law~\eqref{eq:Fadaption} only identifies the forbidden output configurations in domains (II) and (III), the extended suppression law~\eqref{eq:suppcondF} predicts \emph{all} forbidden events (domains (I)-(IV)). Finally, domain (V) collects all finite probability output events for indistinguishable fermions.  We note that the extended suppression law~\eqref{eq:suppcondF} is still formulated as a sufficient condition, and a scenario with suppressed multi-fermion states which are not predicted by~\eqref{eq:suppcondF} is shown in Sec.~\ref{sec:DiscreteFourierTransform} below.

%----------------------------------------------------------------------------------
%            Suppression Law for Bosons
%----------------------------------------------------------------------------------
\subsection{The wave function-based approach}
\label{sec:Bosons}
In the scattering matrix approach outlined in the previous section, we explicitly investigated under which condition permanent and determinant of the scattering matrix $M$ vanish, respectively. Yet, the characteristics of $M$ are inherited from the overlap $\langle\Psi^\mathrm{B/F}(\vec{s})|\Psi_{\mathrm{evo}}^\mathrm{B/F}(\vec{r})\rangle$  (see Eq.~\eqref{eq:transprob}) and, thus, from the properties of the involved many-particle state. We now show that our suppression laws can be consistently derived simply by considering the permutation symmetries of $|\Psi^\mathrm{B/F}(\vec{r})\rangle$. As we proceed to Sec.~\ref{sec:purestates}, this approach will allow us to identify forbidden transitions for even more general states than the so far considered Fock states defined in Eq.~\eqref{eq:state}.

We begin with indistinguishable bosons: The appropriate commutation relations~\eqref{eq:com.bosons} ensure that the state~\eqref{eq:state}, defined by the particle configuration $\vec{r}$, is symmetric under particle exchange. From condition~\eqref{eq:rinvariant}, we further have $r_{\pi(j)}=r_j$, such that 
\begin{align}
\label{eq:Br1}\ket{\Psi^\mathrm{B}(\vec{r})}=&\prod_{j=1}^n\frac{(\opd{a}_j)^{r_j}}{\sqrt{r_j!}}\ket{0}\\
\label{eq:Br2}=&\prod_{j=1}^n\frac{(\opd{a}_{\pi(j)})^{r_j}}{\sqrt{r_j!}}\ket{0}.
\end{align}
Considering the transformation~\eqref{eq:atransform} of creation operators under the action of $U$, Eq.~\eqref{eq:Br1} evolves into 
\begin{align}\label{eq:fin}
\ket{\Psi_{\mathrm{evo}}^\mathrm{B}(\vec{r})}=\prod_{j=1}^n\frac{1}{\sqrt{r_j!}}\left(\sum_{k=1}^nU_{j,k}\opd{b}_k \right)^{r_j}\ket{0}.
\end{align}
On the other hand, for Eq.~\eqref{eq:Br2} we find
\begin{align}
\nonumber\ket{\Psi_{\mathrm{evo}}^\mathrm{B}(\vec{r})}=&\prod_{j=1}^n\frac{1}{\sqrt{r_j!}}\left(\sum_{k=1}^nU_{\pi(j),k}\opd{b}_k \right)^{r_j}\ket{0}\\
\label{eq:fin2}=&\prod_{j=1}^n\frac{1}{\sqrt{r_j!}}\left(\sum_{k=1}^nU_{j,k}\lambda_k\opd{b}_k \right)^{r_j}\ket{0},
\end{align}
where we used the symmetric phase relation~\eqref{eq:sym} and $\prod_{j=1}^n \exp\left(i[\theta(\pi(j))-\theta(j)]\right)=1$. As to be expected, the dependence on $\theta$ cancels out, since it does not affect many-particle interference. 

Let us compare Eq.~\eqref{eq:fin} and~\eqref{eq:fin2}: Obviously, the two expressions only differ by the multiplication of each creation operator $\opd{b}_k$ with the eigenvalue $\lambda_k$, in Eq.~\eqref{eq:fin2}. Forming the overlap $\bracket{\Psi^\mathrm{B}(\vec{s})}{\Psi_{\mathrm{evo}}^\mathrm{B}(\vec{r})}$ with a final state~\eqref{eq:statefinal} defined by the particle configuration $\vec{s}$, Eqs.~\eqref{eq:fin} and~\eqref{eq:fin2} lead to expressions which differ by a factor $\prod_{j=1}^n \lambda_j^{s_j}$. It follows that the overlap must vanish unless $\prod_{j=1}^n \lambda_j^{s_j}=\prod_{\alpha=1}^N \Lambda_\alpha(\vec{s})=1$.  Hence, we obtain for the transition probability in Eq.~\eqref{eq:transprob}:
\begin{align*}
P_\mathrm{B}(\vec{r},\vec{s},U)=0 \hspace{0.5cm} \text{if}  \hspace{0.5cm}\prod_{\alpha=1}^N \Lambda_\alpha(\vec{s})\neq 1,
\end{align*}
which coincides with the suppression law~\eqref{eq:bosonsupplaw} for bosons.

Next, the case of indistinguishable fermions: The adapted suppression law~\eqref{eq:Fadaption} can be derived analogously to the bosonic case, with the difference that, due to the anticommutation relation~\eqref{eq:com.fermions} for fermionic creation operators, the relation between the two versions of the initial state, corresponding to~\eqref{eq:Br1} and \eqref{eq:Br2}, reads
\begin{align}
\nonumber\ket{\Psi^\mathrm{F}(\vec{r})}=&\prod_{j=1}^n (\opd{a}_j)^{r_j}\ket{0}\\
\label{eq:inversions}=&(-1)^w\prod_{j=1}^n(\opd{a}_{\pi(j)})^{r_j}\ket{0},
\end{align} 
with $w$ from Eq.~\eqref{eq:detMsgn}. We then find with the same reasoning as above: 
\begin{align*}
P_\mathrm{F}(\vec{r},\vec{s},U)=0 \hspace{0.5cm} \text{if} \hspace{0.5cm} \prod_{\alpha=1}^N \Lambda_\alpha(\vec{s})\neq (-1)^w.
\end{align*}

Note that, depending on the parity of the initial state permutation as determined by $w$, condition~\eqref{eq:Fadaption} either coincides with the bosonic suppression law~\eqref{eq:bosonsupplaw} (except for a possible multiple occupation of modes in the bosonic case), or the laws are distinct. This was noticed in previous investigations \cite{Tichy-MP-2012,Dittel-MB-2017}, and can now be clearly attributed to the permutation symmetry of the many-particle state. 

The derivation of the extended suppression law for fermions from the input state's permutation symmetry is rather involved and can be found in App.~\ref{appsec:proof2}. Our present wave function-based approach thus leads to results in perfect agreement with the findings of Sec.~\ref{sec:scatteringMatrix}. This is to be expected, since the transition probabilities in Eq.~\eqref{eq:perm} and \eqref{eq:det} result from considerations on the overlap between the initial and final state, and this overlap also was the point of departure for our derivations in the present section. However, while the scattering matrix approach is designed for investigations of many-particle Fock product states on input, the wave function based approach also allows to deal with arbitrary pure input states which are not necessarily separable. This aspect is further elaborated on in Sec.~\ref{sec:purestates}.

%----------------------------------------------------------------------------------
%            Applications of the suppression laws
%----------------------------------------------------------------------------------
\section{Applications of the suppression laws}
\label{sec:Applications}
The generality of our suppression laws manifests itself in the fact that, for each initial particle configuration $\vec{r}$, an entire class of unitary transformation matrices can be determined that exhibit totally destructive interference. In contrast, all previous approaches \cite{Lim-GH-2005,Tichy-ZT-2010,Tichy-MP-2012,Crespi-SL-2015,Dittel-MB-2017,Viggianiello-EG-2017} studied a \emph{given} unitary matrix, to identify input and output configurations which define a suppressed transmission event. In all these cases studied earlier, the input configurations are required to be invariant under certain permutations, in order to infer a suppression law. By means of these specific permutations, we can now verify that all these unitaries indeed exhibit a symmetric phase relation~\eqref{eq:MatrixSym}, and that, therefore, the associated specific suppression laws fall into our present general description of totally destructive many-particle interference. 

%----------------------------------------------------------------------------------
%            Discrete Fourier Transform
%----------------------------------------------------------------------------------
\subsection{Discrete Fourier transform}
\label{sec:DiscreteFourierTransform}
As a first example, we consider the discrete Fourier transform, for which suppression laws were formulated \cite{Lim-GH-2005,Tichy-ZT-2010,Tichy-MP-2012} and tested experimentally \cite{Carolan-UL-2015,Crespi-SL-2016,Su-MI-2017}. The elements of the unitary transformation matrix under consideration read 
\begin{align}\label{eq:UFT}
U_{j,k}^{\mathrm{FT}}=\frac{1}{\sqrt{n}}\exp\left( i \frac{2\pi}{n} (j-1) (k-1)\right)
\end{align}
for all $j,k\in\{1,\dots,n\}$. Initial particle configurations $\vec{r}$ are considered to be periodic with smallest period $\chi$, such that $m=n/\chi\in\mathbb{N}$ is the number of periods, i.e. the length of cycles, and $N/m \in \mathbb{N}$ corresponds to the number of particles per period \cite{Tichy-MP-2012}. Accordingly, we define the permutation 
\begin{align}
\label{eq:DFTsym}
\pi^{\mathrm{FT}}(j)=1+\modu\left[j+\chi-1\ ,\ n\right],
\end{align}
which consists of $\chi$ cycles with length $m$. It follows that the number of distinct cycle lengths is $L=1$.  By construction, the corresponding permutation operator $\mathscr{P}^{\mathrm{FT}}$ leaves $\vec{r}$ invariant, and by plugging Eq.~\eqref{eq:DFTsym} in~\eqref{eq:UFT}, we find the symmetric phase relation of the unitary
\begin{align*}
U^{\mathrm{FT}}_{\pi^{\mathrm{FT}}(j),k}=U^{\mathrm{FT}}_{j,k}\exp\left( i \frac{2\pi}{m} (k-1)\right).
\end{align*} 
Comparing this to the general condition~\eqref{eq:sym}, one can identify the unitary as an eigenbasis of the permutation operator, with the eigenvalues $\lambda_k=\exp\left( i \frac{2\pi}{m} (k-1)\right)$ and $Z=\unit$ in~\eqref{eq:MatrixSym}. The bosonic suppression law~\eqref{eq:bosonsupplaw} predicts the suppression of all output events $\vec{s}$ with
\begin{align}\label{eq:supplawFT}
\prod_{\alpha=1}^N \exp\left({i\frac{2\pi}{m} (d_\alpha(\vec{s})-1)}\right)=\prod_{\alpha=1}^N \exp\left({i\frac{2\pi}{m} d_\alpha(\vec{s})}\right)\neq 1.
\end{align}
With the period $\chi=n/m$ of the initial particle configuration, condition~\eqref{eq:supplawFT} can be rephrased as
\begin{align}\label{eq:supplawFT2}
\modu\left[\chi\sum_{\alpha=1}^N d_\alpha(\vec{s}),n \right]\neq 0,
\end{align}
which coincides with the formulation of \cite{Tichy-ZT-2010,Tichy-MP-2012}. 

Note that we assumed $\chi$ to be the smallest period of the initial state. In general, $\vec{r}$ can also be invariant under permutations $\pi^{\mathrm{FT}}$ with periods $\chi'$ in~\eqref{eq:DFTsym} which are integer multiples of $\chi$. From Eq.~\eqref{eq:supplawFT2} it follows that final configurations $\vec{s}$ which are \emph{allowed} for $\chi$ are likewise allowed for $\chi'$. In turn, if $\vec{s}$ is \emph{suppressed} for any period $\chi'=l \chi$ with $l\in\mathbb{N}$, it is also suppressed for period $\chi$. 

Further, note that single-particle dynamics alone do not lead to any suppression for the Fourier transform, since $U_{j,k}^{\mathrm{FT}}\neq 0$ for all $j,k\in\{1,\dots,n\}$. This is in agreement with condition~\eqref{eq:suppsingle}, given that the permutation $\pi^{\mathrm{FT}}$ only consists of cycles with the same length, that is $L=1$.

Finally, to recover the fermionic suppression law derived in \cite{Tichy-MP-2012}, we apply our above adapted suppression law~\eqref{eq:Fadaption}, for which the number of inversions in Eq.~\eqref{eq:detMsgn} is given by $w=(m-1)N/m$ (note the discussion above Eq.~\eqref{eq:Appw}). Since $(-1)^w=1$, for even $N/m$ or odd $N$, and $(-1)^w=-1$, for odd $N/m$ and even $N$, we obtain
\begin{itemize}
\item for even $N/m$ or odd $N$:
\begin{align} \label{eq:FTichy01} 
\modu\left[\chi \sum_{\alpha=1}^{N}  d_\alpha(\vec{s}) ,n  \right] \neq 0 \Rightarrow P_{\mathrm{F}}(\vec{r},\vec{s},U^{\mathrm{FT}})=0,
\end{align} 
\item for odd $N/m$ and even $N$: 
\begin{align} \label{eq:FTichy02}
\modu\left[\chi \sum_{\alpha=1}^N d_{\alpha}(\vec{s}) , n \right]\neq \frac{n}{2}\Rightarrow P_{\mathrm{F}}(\vec{r},\vec{s},U^{\mathrm{FT}})=0,
\end{align}
\end{itemize}
in agreement with \cite{Tichy-MP-2012}. However, we showed above that our extended suppression law~\eqref{eq:suppcondF} provides an even stronger criterion as compared to~\eqref{eq:Fadaption}, and therefore improves beyond the hitherto known suppression law \cite{Tichy-MP-2012}. We find that all transmission events are suppressed for which each $m$th root of unity does not occur exactly $N/m$ times in the final eigenvalue distribution $\Lambda(\vec{s})$.

Let us illustrate this with the specific example considered in \cite{Tichy-MP-2012}, for $N=4$ indistinguishable fermions injected into $n=12$ modes (see Fig.~4(b) in \cite{Tichy-MP-2012}). For the input state $\vec{d}(\vec{r})=(1,4,7,10)$ (bottom row in the figure), which is invariant under $\pi^{\mathrm{FT}}$ for the smallest period $\chi=3$,  suppressed output events were distinguished as ``predicted''   (black tiles in the figure) or ``unpredicted" (green tiles) by the suppression law of \cite{Tichy-MP-2012}. An unpredicted incident is the output event $\vec{d}(\vec{s})=(1,2,5,6)$: For $\chi=3$, $n=12$ and $m=4$, we have even $N$, odd $N/m$, and Eq.~\eqref{eq:FTichy02} yields $\modu[\chi\sum_{\alpha=1}^N d_\alpha(\vec{s}) ,n]=6=n/2$, and thus does \emph{not} predict this event to be suppressed. In contrast, according to the extended suppression law~\eqref{eq:suppcondF}, we need to compare the final eigenvalue distribution to the initial eigenvalue distribution. For $m=4$, the eigenvalues read $\lambda_k=e^{i\frac{\pi}{2}(k-1)}$, and the initial eigenvalue distribution is $\Lambda_\mathrm{ini}=\{1,-i,-1,i\}$ while the final eigenvalue distribution of $\vec{s}$ is given by $\Lambda(\vec{s})=\{1,i,1,i\}\neq\Lambda_\mathrm{ini}$. This implies that the output event $\vec{s}$ must be suppressed. 

A closer inspection reveals that \emph{all} ``unpredicted" events for this input state as well as for the input state $\vec{d}(\vec{r})=(1,2,7,8)$ are now identified by the extended suppression law~\eqref{eq:suppcondF}. However, the input state $\vec{d}(\vec{r})=(1,3,7,9)$ (second-to-last line in Fig.~4(b) in \cite{Tichy-MP-2012}) exhibits some further suppressed output events, which remain unpredicted, for example the state $\vec{d}(\vec{s})=(1,4,6,7)$. The origin of this effect, as well as for suppressed outputs of non-periodic input states is open for future investigation.

%----------------------------------------------------------------------------------
%            Sylvester matrices and hypercubes
%----------------------------------------------------------------------------------
\subsection{Sylvester matrices and hypercubes}
\label{sec:SylvesterMatricesandHypercubes}
The unitary form of Sylvester matrices $U^{\mathrm{S}}$, and the hypercube unitary $U^{\mathrm{H}}$ for a suitably chosen evolution time can be discussed together, since they only differ by local phase operations ($\Theta$ and $\Sigma$ in Eq.~\eqref{eq:UA}). This is apparent from their respective matrix representations \cite{Crespi-SL-2015,Dittel-MB-2017,Viggianiello-EG-2017}
\begin{align}\label{eq:Sylvester}
U^{\mathrm{S}}=\frac{1}{\sqrt{n}}\begin{pmatrix}1&1\\1&-1\end{pmatrix}^{\otimes d}\end{align}
and
\begin{align}\label{eq:HCmatrix}
U^{\mathrm{H}}=\frac{1}{\sqrt{n}}\begin{pmatrix}1&i\\i&1\end{pmatrix}^{\otimes d},
\end{align}
with the latter obtained from the former upon multiplication of the second row and column of~\eqref{eq:Sylvester} by $i$. Here, $d=\log_2 n\in\mathbb{N}$ is the dimension of the hypercube and Sylvester interferometer. As elaborated in detail in \cite{Dittel-MB-2017}, we assume that the initial particle configuration under consideration is invariant under a permutation $\pi^{\mathrm{SH}}$ with period $m=2$. The permutation $\pi^{\mathrm{SH}}$ is specified by a set $\vec{p}=(p_1,p_2,\dots)$ with $p_i\neq p_j$ for all $i\neq j$ and $p_i\in\{2,4,8,\dots,n\}$: Each element $p_i$ of $\vec{p}$, corresponds to a pairwise exchange of all modes along dimension $\log_2 p_i$ which is obtained in the tensor product form of Eq.~\eqref{eq:Sylvester} and~\eqref{eq:HCmatrix} by a Pauli $\sigma_x$ operation acting on factor $\log_2 p_i$. As detailed in \cite{Dittel-MB-2017}, this permutation reads
\begin{align*}
\pi^{\mathrm{SH}}(j)=j + \sum_{p_k\in \vec{p}}x(j,p_k)\frac{n}{p_k}
\end{align*}
where $x(.,.)$ denotes the Rademacher functions \cite{Rademacher-ES-1922} defined as
\begin{align*}
x(j,p)=(-1)^{\lfloor\frac{p(j-1)}{n}\rfloor},
\end{align*}
with the floor function $\lfloor z\rfloor$ rendering the greatest integer that is less than, or equal to $z$. By utilizing the Walsh functions \cite{Walsh-CS-1923}
\begin{align*}
\mathcal{A}(j,\vec{p})=\prod_{p_k\in\vec{p}}x(j,p_k)
\end{align*}
one obtains the symmetric phase relations~\eqref{eq:sym} which read, for all $j,k\in\{1,\dots,n\}$:
\begin{align*}
U^{\mathrm{S}}_{\pi^{\mathrm{SH}}(j),k}=U^{\mathrm{S}}_{j,k}\exp\left( i\pi\left[\frac{\mathcal{A}(k,\vec{p})-1}{2}\right] \right)
\end{align*}
and
\begin{align}\label{eq:HCsymrel}
U^{\mathrm{H}}_{\pi^{\mathrm{SH}}(j),k}=&U^{\mathrm{H}}_{j,k}\exp\left( i\pi \left[\frac{\mathcal{A}(k,\vec{p})-1}{2}\right]\right)\\
\nonumber&\times\exp\left(i\left[\theta^\mathrm{H}(\pi(j))-\theta^\mathrm{H}(j)\right]\right)
\end{align}
where $\theta^\mathrm{H}(j)=\frac{\pi}{4}\sum_{l=1}^{d}[1-x(j,2^l)]$. Note that the equivalence of Eq.~(\ref{eq:HCsymrel}) with the phase relation given in \cite{Dittel-MB-2017} can be proven by induction. Since the similarity of both unitaries is established by
\begin{align*}
U^{\mathrm{H}}=\Theta^\mathrm{H}U^{\mathrm{S}}\Sigma^\mathrm{H},
\end{align*}
with $\Theta^\mathrm{H}_{j,j}=\Sigma^\mathrm{H}_{j,j}=\exp(i\theta^\mathrm{H}(j))$, suppressed transmission events are the same in both cases and can be identified by inspection of the eigenvalues $\lambda_k=\exp(i\pi\frac{\mathcal{A}(d_k,\vec{p})-1}{2})$. 

Exploiting our bosonic suppression law~\eqref{eq:bosonsupplaw}, we find that for an input state invariant under $\pi^{\mathrm{SH}}$, all those output events $\vec{s}$ are suppressed, for which
\begin{align}\label{eq:HCsuppcond}
\prod_{\alpha=1}^{N} \exp\left(i\pi \frac{\mathcal{A}(d_\alpha(\vec{s}),\vec{p})-1}{2}\right) \neq 1.
\end{align}
With $\mathcal{A}(.,.)\in\{1,-1\}$, this is equivalent to
\begin{align*}
\prod_{\alpha=1}^{N} \mathcal{A}(d_\alpha(\vec{s}),\vec{p})=-1\ \Rightarrow\ P_{\mathrm{B}}(\vec{r},\vec{s},U^{\mathrm{S,H}})=0,
\end{align*}
which coincides with the findings in \cite{Dittel-MB-2017}. 

For fermions, by virtue of~\eqref{eq:suppcondF}, output events $\vec{s}$ with $\Lambda(\vec{s})\neq\Lambda_\mathrm{ini}$ are suppressed. Since $\Lambda_\mathrm{ini}$ contains each eigenvalue $1$ and $-1$ exactly $N/2$ times, we find the condition for suppression as stated in \cite{Dittel-MB-2017}:
\begin{align*}
\sum_{\alpha=1}^{N}\mathcal{A}(d_\alpha(\vec{s}),\vec{p})\neq 0\ \Rightarrow\ P_{\mathrm{F}}(\vec{r},\vec{s},U^{\mathrm{S,H}})=0.
\end{align*}

As a final note, just as for the Fourier matrix considered in Sec.~\ref{sec:DiscreteFourierTransform}, $U^\mathrm{S,H}_{j,k}\neq 0$ for all $j,k\in\{1,\dots,n\}$. Therefore, single-particle dynamics cannot explain any of the suppressed events under the action of $U^\mathrm{S,H}$.

%----------------------------------------------------------------------------------
%            $J_x$ Unitary
%----------------------------------------------------------------------------------
\subsection{$J_x$ Unitary}
\label{sec:JxUnitary}
In our last example, we focus on the $J_x$ unitary, for which suppression laws were formulated for $N=2$ particles, and experimentally verified with bosons \cite{Perez-Leija-CQ-2013,Weimann-IQ-2016}. By means of the above considerations, we now show that this unitary encodes the eigenbasis of a certain permutation operation, except for local phase operations. This then allows us to generalise the hitherto known results to bosonic and fermionic configurations, with arbitrary particle number. 

The matrix representation of the $J_x$ unitary is generated by the angular momentum operator in $x$-direction, $U^{\mathrm{J}}(t)=e^{iJ_xt/\hbar}$, with \cite{Perez-Leija-CQ-2013,Perez-Leija-PT-2013,Weimann-IQ-2016}
\begin{align*}
\left[J_x\right]_{j,k}=\frac{\hbar}{2}\left(\sqrt{k(n-k)}\delta_{j,k+1}+\sqrt{j(n-j)}\delta_{j,k-1} \right).
\end{align*}
For an evolution time $t=\pi/2$, the unitary can be expressed in terms of its own eigenstates $\ket{u^{(j)}}$. With the notation $U^{\mathrm{J}}(\pi/2)\equiv U^{\mathrm{J}}$, its elements read \cite{Perez-Leija-PT-2013,Weimann-IQ-2016}
\begin{align*}
U^{\mathrm{J}}_{j,k}=e^{i\frac{\pi}{2}(j-k)}u^{(j)}_k,
\end{align*}
where $u^{(j)}_k$ denotes the $k$th component of eigenstate $\ket{u^{(j)}}$ and is given by \cite{Perez-Leija-CQ-2013,Perez-Leija-PT-2013,Weimann-IQ-2016}
\begin{align*}
u^{(j)}_k=2^{-\frac{1}{2}(n+1)+k}\sqrt{\frac{(k-1)!(n-k)!}{(j-1)!(n-j)!}}\ P_{k-1}^{(j-k,n-j-k+1)}(0)
\end{align*}
with $P_{\gamma}^{(\alpha,\beta)}(0)$ the Jacobi polynomial \cite{Abramowitz-HM-1964} of order $\gamma$ evaluated at the origin. By means of the symmetries proper to the Jacobi polynomials, one finds
\begin{align}\label{eq:UJxsym1}
U^{\mathrm{J}}_{n+1-j,k}=U^{\mathrm{J}}_{j,k} \exp\left(i\pi\left[k-j+\frac{n-1}{2}\right]\right).
\end{align}
It thus becomes evident that a mirror symmetry with respect to the central mode governs the dynamics. The pertinent permutation $\pi^{\mathrm{J}}$ with period  $m=2$ reads
\begin{align*}
\pi^{\mathrm{J}}(j)=n+1-j,
\end{align*}
for all $j\in\{1\dots,n\}$. Note that, for odd $n$, the mode $j=(n+1)/2$ is unaffected by $\pi^{\mathrm{J}}$, that is $\pi^{\mathrm{J}}((n+1)/2)=(n+1)/2$. Consequently, the permutation $\pi^{\mathrm{J}}$ consists of $(n-1)/2$ cycles of length $2$, and of one cycle of length $1$, such that $L=2$. For even $n$, however, $L=1$, since there are only cycles of length $2$. 

With the above, we can characterise the suppressed events in the general scenario, for arbitrary $n$ and $N$. By~\eqref{eq:UJxsym1}, the unitary's symmetric phase relation~\eqref{eq:sym} reads, for all $j,k\in\{1,\dots,n\}$, 
\begin{align*}
\begin{split}
U^{\mathrm{J}}_{\pi^{\mathrm{J}}(j),k}=&U^{\mathrm{J}}_{j,k}\exp\left(i\pi \left(k-1\right)\right)\\
\times&\exp\left(i\left[\theta^\mathrm{J}(\pi^{\mathrm{J}}(j))-\theta^\mathrm{J}(j)\right]\right),
\end{split}
\end{align*}
with $\theta^\mathrm{J}(j)=\pi j/2$. This identifies the eigenvalues $\lambda_k=\exp(i\pi(k-1))$ which we need to formulate suppression laws. For bosonic input configurations which are invariant under $\pi^{\mathrm{J}}$ (i.e., mirror-symmetric), our suppression law reveals that all output events which adhere to
\begin{align*}
\prod_{\alpha=1}^{N} \exp\left(i\pi(d_\alpha(\vec{s})-1)\right) \neq 1
\end{align*}
must be suppressed. Accordingly, all events with an odd number of bosons transmitted into modes with even label (corresponding to negative eigenvalues) are suppressed. The results obtained for the special case $N=2$ \cite{Perez-Leija-CQ-2013,Weimann-IQ-2016} follow directly from this more general condition.

For fermions, our suppression law~\eqref{eq:suppcondF}, together with the fact that $1$ and $-1$ are the only eigenvalues of the permutation operator of $\pi^{\mathrm{J}}$, implies:
\begin{itemize}
\item for odd $n$ and odd $N$,
\begin{align}\label{eq:JxFc1}
\abs{\{\lambda\in\Lambda(\vec{s}) : \lambda=1\}}\neq\frac{N+1}{2}\ \Rightarrow\ P_\mathrm{F}(\vec{r},\vec{s},U^{\mathrm{J}})=0,
\end{align}
\item otherwise,
\begin{align*}
\abs{\{\lambda\in\Lambda(\vec{s}) : \lambda=1\}}\neq\frac{N}{2}\ \Rightarrow\ P_\mathrm{F}(\vec{r},\vec{s},U^{\mathrm{J}})=0,
\end{align*}
\end{itemize}
which includes also the special case $N=2$ \cite{Perez-Leija-CQ-2013}.

The $J_x$ unitary with an odd number of modes also offers an example where single-particle dynamics enforce forbidden many-particle transmission events. As discussed above, for odd $n$, the permutation $\pi^\mathrm{J}$ consists of one cycle with length $m_1=1$, and of other cycles all with length $m_2=2$. In the following, we assume that there is at least one particle injected into mode $(n+1)/2$, which corresponds to the cycle with length $1$, that is $N_1>0$. Following our discussion in Sec.~\ref{sec:zerotransmission}, \eqref{eq:suppsingle} reveals that all many-particle configurations for which
\begin{align}\label{eq:JxTcond1}
\mathscr{N}_1(\vec{s})<N_1
\end{align}
or
\begin{align}\label{eq:JxTcond2}
\mathscr{N}_2(\vec{s})<N_2
\end{align}
are forbidden, irrespective of the particles' mutual distinguishability. In our present case, $\mathscr{N}_2(\vec{s})=N$ by \eqref{eq:Ndef}, and condition~\eqref{eq:JxTcond2} can never be fulfilled. This leaves us with condition~\eqref{eq:JxTcond1}, which states that all the $N_1$ particles starting in the central mode must exit in an odd mode (associated with eigenvalue $1$), or, equivalently, that all allowed transmission events can transmit at most $N_2$ particles into even modes. 

In the case of indistinguishable fermions, condition~\eqref{eq:JxTcond1} is already built into the fermionic suppression law~\eqref{eq:JxFc1}: Since we require odd $n$ and $N_1>0$, we have $N_1=1$ while $N_2=N-1$ must be even, so that $N$ is odd. By~\eqref{eq:Ndef}, output events are forbidden for which $\mathscr{N}_1(\vec{s})=\abs{\{\lambda\in\Lambda(\vec{s}) : \lambda=1\}}<N_1=1$. Accordingly, $\abs{\{\lambda\in\Lambda(\vec{s}) : \lambda=1\}}=0$ for these events, which are already covered by the suppression law~\eqref{eq:JxFc1} since $0\neq (N+1)/2$.

In conclusion, all hitherto known unitaries which exhibit totally destructive many-particle interference are retrieved within our framework: Up to local phase operations $\Theta$ and $\Sigma$ (recall Eq.~\eqref{eq:UA}), these unitaries diagonalise the permutation operator which leaves the initial state invariant. By rotations in the degenerate subspaces (recall the discussion at the end of Sec.~\ref{sec:Unitarymatrices}), they can be related to their block-diagonal canonical matrices $A_\mathrm{C}$ consisting of Fourier-unitaries which diagonalise the individual cycles of the permutation. It is worth noting that some unitaries (up to local phase operations) can diagonalise multiple permutation operators and, thus, give rise to multiple suppression laws. One of these is the hypercube unitary~\eqref{eq:HCmatrix} which simultaneously diagonalises permutation operators corresponding to different sets $\vec{p}$. For initial configurations invariant under several of such permutations, this leads to rich suppression effects as detailed in \cite{Dittel-MB-2017}.

%----------------------------------------------------------------------------------
%            Suppression for arbitrary Pure States
%----------------------------------------------------------------------------------
\section{Suppression for arbitrary Pure States}\label{sec:purestates}
In the previous sections, our discussion was restricted to initial states of the form~\eqref{eq:state}, that is, we only considered configurations $\vec{r}$ of indistinguishable particles with a definite number of particles per mode. Now we relax this assumption and consider any initial pure state $\ket{\Psi\{\opd{a}_{j,\ket{I}}\}}$ which can, in some way, be expressed by creation operators $\opd{a}_{j,\ket{I}}$ acting on the vacuum $\ket{0}$. Note that this does not assume a maximal total number of particles and not even a fixed total particle number.  While $j$ again labels the mode number $1,\dots,n$, we moreover consider additional degrees of freedom of the particles, which are specified by the \textit{internal states} $\ket{I}$. Unless otherwise stated, we assume that different internal states are not necessarily orthogonal to each other. In our approach, we jointly treat bosons and fermions, keeping in mind the (anti)commutation relations~\eqref{eq:com.bosons} and~\eqref{eq:com.fermions}. 

Much as above, we start out from an arbitrary permutation operation $\mathscr{P}$ that leaves $\ket{\Psi\{\opd{a}_{j,\ket{I}}\}}$ unchanged except for a phase $\varphi\in\mathbb{R}$:
\begin{align}\label{eq:PsiIP}
\ket{\Psi\{\opd{a}_{j,\ket{I}}\}}\xrightarrow[]{\mathscr{P}} e^{i\varphi}\ket{\Psi\{\opd{a}_{j,\ket{I}}\}}.
\end{align}
As before, $\mathscr{P}$ shall only permute modes $j\in\{1,\dots,n\}$, that is
\begin{align}\label{eq:aIP}
\opd{a}_{j,\ket{I}}\xrightarrow[]{\mathscr{P}}\opd{a}_{\pi(j),\ket{I}},
\end{align}
while leaving the internal state unaffected. In a unitary evolution, interference effects exhibited by states of the form $\ket{\Psi\{\opd{a}_{j,\ket{I}}\}}$ are generally sensitive to local phase operations on the input modes. We therefore assume that there are no initially imprinted local phases and set $\Theta=\unit$ in Eq.~\eqref{eq:UA}. The evolution of $\ket{\Psi\{\opd{a}_{j,\ket{I}}\}}$ is then described by the unitary
\begin{align}\label{eq:UASigma}
U=A~\Sigma,
\end{align}
with -- as before -- $A$ being any eigenbasis of $\mathscr{P}$, and $\Sigma$ accounting for arbitrary local phase operations on final modes. Since we exclude initial phase operations, the symmetric phase relation~\eqref{eq:sym} simplifies to $U_{\pi(j),k}=U_{j,k}\lambda_k$, such that the evolution of the creation operators $\opd{a}_{j,\ket{I}}$ and $\opd{a}_{\pi(j),\ket{I}}$ can be expressed as follows:
\begin{align}
\opd{a}_{j,\ket{I}}&\xrightarrow[]{U}\sum_{k=1}^n U_{j,k} \opd{b}_{k,\ket{I}}\label{eq:aI1}\\
\opd{a}_{\pi(j),\ket{I}}&\xrightarrow[]{U}\sum_{k=1}^n U_{j,k}\lambda_k\opd{b}_{k,\ket{I}}.\label{eq:aI2}
\end{align}

As above, we are interested in the suppression of specific output mode occupations. Therefore, we investigate whether the evolved state $\ket{\Psi_{\mathrm{evo}}\{\opd{a}_{j,\ket{I}}\}}$ -- the image of $\ket{\Psi\{\opd{a}_{j,\ket{I}}\}}$ under~\eqref{eq:aI1} -- has non-vanishing overlap with any of the final states
\begin{align*}
\ket{\Psi(\vec{s},\Omega)}\propto\prod_{\alpha=1}^N\opd{b}_{d_\alpha(\vec{s}),\ket{I_\alpha}}\ket{0},
\end{align*}
which represent sharp output mode occupations, with an arbitrary set $\Omega=(\ket{I_1},\dots,\ket{I_N})$ of internal states $\ket{I_\alpha}$ in the final mode $d_\alpha(\vec{s})$. In particular, we examine under which conditions $\langle\Psi(\vec{s},\Omega)|\Psi_{\mathrm{evo}}\{\opd{a}_{j,\ket{I}}\}\rangle=0$ for all possible $\Omega$.

According to~\eqref{eq:PsiIP} and~\eqref{eq:aIP}, we have 
\begin{align}\label{eq:phasePi}
\ket{\Psi\{\opd{a}_{j,\ket{I}}\}}=e^{-i\varphi}\ket{\Psi\{\opd{a}_{\pi(j),\ket{I}}\}}.
\end{align}
In close analogy to the comparison of Eq.~\eqref{eq:fin} with~\eqref{eq:fin2} for product states, one finds, by comparison of Eq.~\eqref{eq:aI1} with~\eqref{eq:aI2}, that the overlap $\langle\Psi(\vec{s},\Omega)|\Psi_{\mathrm{evo}}\{\opd{a}_{j,\ket{I}}\}\rangle$ must be zero for all $\Omega$ unless $e^{-i\varphi}\prod_{\alpha=1}^N\lambda_{d_\alpha(\vec{s})}=1$, since each creation operator $\opd{b}_{k,\ket{I}}$ in Eq.~\eqref{eq:aI2} is accompanied by the eigenvalue $\lambda_k$, irrespective of $\ket{I}$. Thus, all final particle configurations $\vec{s}$ must be suppressed for which
\begin{align}\label{eq:suppcondgeneral}
\prod_{\alpha=1}^N\lambda_{d_\alpha(\vec{s})}=\prod_{\alpha=1}^N\Lambda_\alpha(\vec{s})\neq e^{i\varphi}.
\end{align}
Note that many-particle Fock product states of the form~\eqref{eq:state} also satisfy the initial assumption in Eq.~\eqref{eq:PsiIP}, with $e^{i\varphi}=1$ for bosons and $e^{i\varphi}=(-1)^w$ for fermions. Equation~\eqref{eq:suppcondgeneral} thus reproduces the bosonic suppression law~\eqref{eq:bosonsupplaw} and the adapted suppression law for fermions~\eqref{eq:Fadaption}, respectively. 

For the general fermionic suppression law~\eqref{eq:suppcondF} we have not been able to derive a similar generalisation for arbitrary pure states. Its derivation shown in App.~\ref{appsec:proof2} reveals no clue towards an adaptation to general pure states $\ket{\Psi\{\opd{a}_{j,\ket{I}}\}}$, and we only find that certain many-particle states such as superpositions of Fock product states which generate the same initial eigenvalue distribution lead to the fermionic suppression condition~\eqref{eq:suppcondF}. This seems natural since the suppression law~\eqref{eq:suppcondF} specifically requires a known initial eigenvalue distribution. However, for arbitrary initial pure state $\ket{\Psi\{\opd{a}_{j,\ket{I}}\}}$, such an initial eigenvalue distribution is not well-defined, because the symmetry~\eqref{eq:PsiIP} may hold for superpositions of product states which generate different initial eigenvalue distributions or even represent different particle numbers. 

Let us stress once again that condition~\eqref{eq:suppcondgeneral} is based on the input state's permutation symmetry~\eqref{eq:PsiIP} alone, regardless of the specific particle type and on the particles' mutual distinguishability. This provides us with a remarkable insight into the connection between many-particle interference and  distinguishability. However, before we pursue this direction further, we discuss two short examples that highlight the general applicability of our formalism to symmetric states of the form~\eqref{eq:PsiIP}.
 
%----------------------------------------------------------------------------------
%            Superpositions of indistinguishable particles
%----------------------------------------------------------------------------------
\subsection{Superpositions of indistinguishable particles}
In our first example, we assume a fixed number $N$ of perfectly indistinguishable bosons such that the internal state $\ket{I}$ is the same for all $N$ bosons under consideration. Further, let $\mathscr{P}$ be any permutation operator corresponding to a permutation $\pi\in\mathrm{S}_n$ of period $m$. For any initial many-particle configuration $\vec{r}$, we create an initial state $\ket{\Phi^\mathrm{B}(\vec{r})}$ obeying Eq.~\eqref{eq:PsiIP} by superposition of the Fock product states defined by $\vec{r},\mathscr{P}\vec{r},\dots,\mathscr{P}^{m-1}\vec{r}$,
\begin{align}\label{eq:statesuperpos}
\ket{\Phi^\mathrm{B}(\vec{r})}=\frac{1}{\sqrt{m}}\sum_{l=0}^{m-1}e^{i\frac{2\pi}{m}lk}\ket{\Psi^\mathrm{B}(\mathscr{P}^l\vec{r})},
\end{align}
with $\ket{\Psi^\mathrm{B}(\vec{r})}$ from Eq.~\eqref{eq:state} and $k$ an arbitrary integer. With $\mathscr{P}^m=\unit$, it is straightforward to verify that 
\begin{align*}
\ket{\Phi^\mathrm{B}(\vec{r})}\xrightarrow[]{\mathscr{P}} e^{-i\frac{2\pi}{m}k}\ket{\Phi^\mathrm{B}(\vec{r})}.
\end{align*}
By~\eqref{eq:PsiIP}, the evolution of $\ket{\Phi^\mathrm{B}(\vec{r})}$ under a unitary~\eqref{eq:UASigma}, suppresses all output events $\vec{s}$ with
\begin{align}\label{eq:supersupp}
\prod_{\alpha=1}^N\Lambda_\alpha(\vec{s})\neq e^{-i\frac{2\pi}{m}k}.
\end{align}

The simplest scenario for such an evolution is the \textit{single-particle router}: Assume we can prepare a single particle in the superposition of $n=m$ modes $\frac{1}{\sqrt{m}}\sum_{l=0}^{m-1} \opd{a}_{l+1}\ket{0}$ (e.g. by performing the Fourier transform $U^\mathrm{FT}$ on $\opd{a}_1 \ket{0}$), and we have control over the phase in each mode, respectively, and set $\ket{\Phi}=\frac{1}{\sqrt{m}}\sum_{l=0}^{m-1}e^{i\frac{2\pi}{m}lk}\opd{a}_{l+1}\ket{0}$ with $k\in \{0,...,m-1\}$. When $\ket{\Phi}$ is imaged through another Fourier transform (which is the canonical unitary matrix associated with $\mathscr{P}$ and satisfies~\eqref{eq:UASigma}, the condition for suppression~\eqref{eq:supersupp} reveals that the particle will definitely be transmitted to the final mode that corresponds to the eigenvalue $e^{-i\frac{2\pi}{m}k}$. Since for each $k\in\{0,\dots,m-1\}$, there is only one such mode, we can route the particle in any desired mode by setting $k$ accordingly. This concept is widely used in optical wavelength-division multiplexing via the insertion of controllable phases in the Fourier plane of multi-channel imaging devices \cite{Dobrusin-DF-2005}.

%----------------------------------------------------------------------------------
%            Suppression for entangled states
%----------------------------------------------------------------------------------
\subsection{Suppression for entangled states}\label{sec:Bell}
In our second example, we include additional degrees of freedom and consider the evolution of an $N$-particle entangled state
\begin{align}\label{eq:entstate}
\ket{\Psi_k}\propto\sum_{l=0}^{m-1}e^{i\frac{2\pi}{m}lk}\prod_{\alpha=1}^N \opd{a}_{\pi^l(d_\alpha(\vec{r})),\ket{I_\alpha}}\ket{0},
\end{align}
with particles in modes $\vec{d}(\vec{r})$, $\vec{r}$ a particle occupation which is invariant under the permutation $\pi$ of order $m$, and the assignment of the particles' internal states varying in each summand. Here $\ket{I_\alpha}$ is the internal state of the $\alpha$th particle, and, if $\bracket{I_\alpha}{I_\beta}=\delta_{\alpha,\beta}$, the missing  prefactor in~\eqref{eq:entstate} is given by $1/\sqrt{m}$. For any $k\in\mathbb{Z}$, the permutation characteristics of these states is given by
\begin{align*}
\ket{\Psi_k}\xrightarrow[]{\mathscr{P}}e^{-i\frac{2\pi}{m}k}\ket{\Psi_k}.
\end{align*}
According to Eqs.~\eqref{eq:PsiIP} and \eqref{eq:suppcondgeneral}, all output events $\vec{s}$ with
\begin{align*}
\prod_{\alpha=1}^N\Lambda_\alpha(\vec{s})\neq e^{-i\frac{2\pi}{m}k}
\end{align*}
are suppressed. For states of the form~\eqref{eq:entstate}, the same suppression behaviour applies as for superpositions~\eqref{eq:statesuperpos} of indistinguishable particles. This highlights that the suppression hinges on the wave function's symmetry -- which is the same in both cases.

A special realization of entangled states \`a la~\eqref{eq:entstate} are the (maximally entangled) Bell states $\ket{\Psi^\pm}$. In this case, we consider the permutation $\pi=(1~2)$ and the mode assignment list $\vec{d}(\vec{r})=(1,2)$ of $N=2$ particles with internal states $\ket{I_1}=\ket{\uparrow}$ and $\ket{I_2}=\ket{\downarrow}$. Using bosonic creation operators and the notation $\opd{a}_{d_\alpha(\vec{r}),\ket{I_\alpha}}\ket{0}=\ket{I_\alpha}_{d_\alpha(\vec{r})}$, Eq.~\eqref{eq:entstate} generates (up to a proportionality factor) the Bell states 
\begin{align}\label{eq:Bellstates}
\ket{\Psi^\pm}\propto\left(\ket{\uparrow}_1\ket{\downarrow}_2 \pm \ket{\downarrow}_1\ket{\uparrow}_2\right)
\end{align}
where $``+"$ corresponds to $k=0$ and $``-"$ to $k=1$. As the permutation $\pi$ performs an exchange of mode $1$ and $2$, we find
\begin{align}\label{eq:Bellsym}
\ket{\Psi^\pm}\xrightarrow[]{\mathscr{P}}\pm\ket{\Psi^\pm}.
\end{align} 
Now consider the evolution of $\ket{\Psi^\pm}\rightarrow\ket{\Psi^\pm_\mathrm{evo}}$ according to the one-dimensional Sylvester matrix~\eqref{eq:Sylvester},
\begin{align*}
U^\mathrm{S}=\frac{1}{\sqrt{2}}\begin{pmatrix}
1&1\\
1&-1
\end{pmatrix},
\end{align*}
which describes an eigenbasis of $\mathscr{P}$ with eigenvalues $\lambda_1=1$ and $\lambda_2=-1$ (and represents the action of a balanced two-mode coupler or beam splitter). According to~\eqref{eq:suppcondgeneral}, all output events $\vec{s}$ for which 
\begin{align}\label{eq:supplawBell}
\lambda_{d_1(\vec{s})}\lambda_{d_2(\vec{s})}\neq \pm 1
\end{align}
must be forbidden. Since the only eigenvalues are $\lambda_1=1$ and $\lambda_2=-1$, we find that $\ket{\Psi^+_\mathrm{evo}}$ exhibits vanishing amplitude for both particles in separate modes, while $\ket{\Psi^-_\mathrm{evo}}$ has no overlap with any state where both particles occupy the same mode, mimicing the behaviour of indistinguishable bosons and fermions, respectively. This has been verified experimentally \cite{Michler-IB-1996,Mattle-DC-1996} and is in agreement with a full calculation which produces
\begin{align*}
\ket{\Psi^+_\mathrm{evo}}&\propto\left(\ket{\uparrow}_1\ket{\downarrow}_1-\ket{\uparrow}_2\ket{\downarrow}_2 \right),\\
\ket{\Psi^-_\mathrm{evo}}&\propto\left(\ket{\downarrow}_1\ket{\uparrow}_2-\ket{\uparrow}_1\ket{\downarrow}_2 \right).
\end{align*}
Here it is worth noting that the suppression law~\eqref{eq:supplawBell} does not require orthogonal internal degrees of freedom $\bracket{\uparrow}{\downarrow}=0$. In fact, any other internal states $\ket{I_1}$ and $\ket{I_2}$ could have been chosen as long as the symmetry~\eqref{eq:Bellsym} is fulfilled.

%----------------------------------------------------------------------------------
%            Total Destructive Interference and Partial Distinguishability
%----------------------------------------------------------------------------------
\subsection{Perfect suppression for partial distinguishability}\label{sec:partialdist}
In the previous section we discussed suppression laws for entangled states without explicit assumptions on the particles' mutual distinguishability. In the case of Bell states $\ket{\Psi^\pm}$, for example, the suppression of the discussed final particle configurations appears independently of the overlap $\bracket{\uparrow}{\downarrow}$. Thus, totally destructive interference is not necessarily affected by mutual particle distinguishability. A similar situation was noticed \cite{Ou-OF-1999,Dufour-MP-2017} for Fock-product states injected into a two-mode coupler and for a particular many-particle Fock product state subject to the discrete Fourier transform~\eqref{eq:UFT}  \cite{Tichy-SP-2015}. A more detailed analysis of the latter case \cite{Shchesnovich-PI-2015} led to the \textit{zero-probability conjecture} that attributes totally destructive interference to an exact cancellation of many-particle amplitudes arising from a subset of completely indistinguishable particles (these particles being partially distinguishable from the remaining particles involved). Moreover, it was conjectured that the suppression persists if the degree of distinguishability between this subset and the other particles is changed. In view of the above scattering scenarios, we now show that the dependence of the described suppression on mutual particle distinguishability is in perfect agreement with the zero-probability conjecture and we further address the origin of this effect.

\begin{figure}[t]
\centering
\includegraphics[width=0.47\textwidth]{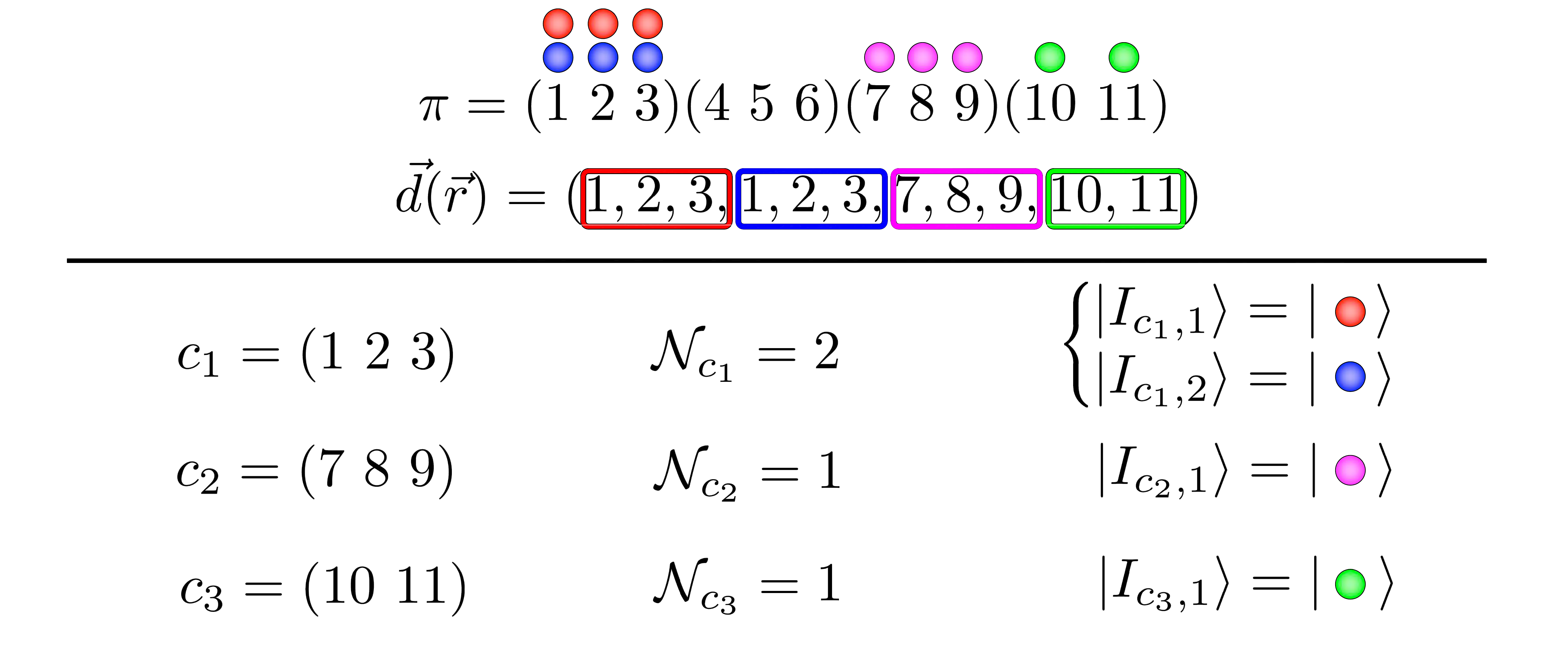} 
\caption{(Color online) Example for a state $\ket{\Psi^\mathrm{B}}$ that is invariant under the permutation $\pi$, but describes many partially distinguishable particles. The particles, illustrated by balls, occupy three different cycles of $\pi$, $c_1=(1~2~3)$, $c_2=(7~8~9)$ and $c_3=(10~11)$. Each mode in cycle $c_k$ is occupied by $\mathcal{N}_{c_k}$ particles. The particles' coloring represent their internal state, given by $\ket{I_{c_k,j}}$ for $k\in\{1,\dots,3\}$ and $j\in\{1,\dots,\mathcal{N}_{c_k}\}$.}
\label{fig:partdist}
\end{figure}

We begin with many-particle product states as in  Eq.~\eqref{eq:state} and investigate under which conditions the attribution of internal states $\ket{I}$ to the particles does not affect the state's permutation symmetry, thereby leaving the suppression law unaltered. For the many-particle state to be invariant under the permutation operator $\mathscr{P}$ associated with $\pi\in\mathrm{S}_n$, all modes $j$ belonging to the same cycle $c$ of $\pi$ must have an identical particle content. As illustrated for an example in Fig.~\ref{fig:partdist}, they must contain the same number $\mathcal{N}_c$ of particles, with the same set of internal states $\{\ket{I_{c,q}}\}_{q=1,\dots,\mathcal{N}_c}$. The initial state can therefore be written as 
\begin{align}\label{eq:ParDistState}
\ket{\Psi^\mathrm{B/F}}\propto \prod_{c\in\mathrm{cycles}(\pi)} \prod_{j\in c} \prod_{q=1}^{\mathcal{N}_c} \opd{a}_{j,\ket{I_{c,q}}} \ket{0}.
\end{align}
The particles are thus divided in sets, labelled by $c\in\mathrm{cycles}(\pi)$ and $q=1,\dots,\mathcal{N}_c$, which all share the same internal state $\ket{I_{c,q}}$. State~\eqref{eq:ParDistState} clearly satisfies~\eqref{eq:PsiIP}, irrespective of the mutual distiguishability between distinct sets of indistinguishable particles. However, once the degree of distinguishability between particles within these sets changes, the initial state violates~\eqref{eq:PsiIP} and the suppression law~\eqref{eq:suppcondgeneral} loses its validity. This exactly coincides with the conjecture in \cite{Shchesnovich-PI-2015}, which is grounded on a decomposition of transition amplitudes corresponding to different sets of indistinguishable particles, just as considered here. Consequently, in our case, the suppression of many-particle product states does not depend on the mutual distinguishability between particles in modes corresponding to different cycles of $\pi$, or particles within a cycle that belong to different sets. Note that this also applies to the extended fermionic suppression law~\eqref{eq:suppcondF} for many-particle Fock product states, which can be verified by including internal states in the derivation in App.~\ref{appsec:proof2}.

For arbitrary pure states $\ket{\Psi\{\opd{a}_{j,\ket{I}}\}}$, particle distinguishability plays a secondary role for totally destructive interference. For example, in Sec.~\ref{sec:Bell} we discussed the suppression for entangled states~\eqref{eq:entstate}. While the state of each individual particle is undefined, the degree of their internal states' mutual indistinguishability can be changed without affecting the suppression of final particle configurations. The suppression arises from the state's symmetry~\eqref{eq:PsiIP} and is unaffected by the particles' mutual indistinguishability, just as in the case of many-particle product states. Thus, it is more relevant to consider the permutation characteristics of the states rather than the property of the particles when making statements on suppressed transmission events: Any changes of the state $\ket{\Psi\{\opd{a}_{j,\ket{I}}\}}$, be it in the mode occupation or in the internal degrees of freedom, for which the permutation symmetry~\eqref{eq:PsiIP} remains unaffected, have no effect on the level of suppressed transmission events as predicted by~\eqref{eq:suppcondgeneral}.

%----------------------------------------------------------------------------------
%            Discussion and Conclusion
%----------------------------------------------------------------------------------
\section{Discussion and Conclusion}
In Sec.~\ref{sec:supplaw} and~\ref{sec:Applications}, we discussed multi-mode scattering of permutation symmetric Fock product states of indistinguishable particles, and found a generic class of unitary transformation matrices which generate vanishing output events, due to perfect destructive interference of many-particle amplitudes. By means of our suppression laws~\eqref{eq:bosonsupplaw}, \eqref{eq:Fadaption} and~\eqref{eq:suppcondF}, we rederived all previously known scattering scenarios that exhibit totally destructive many-particle interference. Moreover, our extended fermionic suppression law~\eqref{eq:suppcondF} exceeds the one hitherto known for the discrete Fourier transform \cite{Lim-GH-2005,Tichy-ZT-2010,Tichy-MP-2012}, and we generalised the earlier suppression law for two particles in the $J_x$ unitary \cite{Perez-Leija-CQ-2013,Weimann-IQ-2016} to arbitrary particle numbers.

Our investigations in Sec.~\ref{sec:purestates} show that the suppression of transmission events does not necessarily require fully indistinguishable particles. Instead, we identified the \emph{many-particle input state's permutation symmetry} as the crucial factor. As highlighted in Sec.~\ref{sec:partialdist}, the validity of our suppression laws, \eqref{eq:bosonsupplaw}, \eqref{eq:Fadaption} and~\eqref{eq:suppcondF}, is thus independent of the mutual particle distinguishability, as long as the permutation symmetry of the initial state remains unaffected.

We found the dependency of the suppression on the permutation symmetry using the wavefunction based approach, introduced in Sec.~\ref{sec:Bosons} for Fock product states, and generalised in Sec.~\ref{sec:purestates} for arbitrary pure states. In the general case, we considered any pure state that is invariant under a permutation operation except for a global phase. This phase then determines the condition for suppressed output events. Naturally, the bosonic~\eqref{eq:bosonsupplaw} and the adapted fermionic suppression law~\eqref{eq:Fadaption}, which are only based on the wave function's permutation symmetry, are contained in this more general approach. However, the extended fermionic suppression law~\eqref{eq:suppcondF} for many-particle \emph{Fock product states} stands by itself. No indication for a generalisation to permutation-symmetric pure states was found, since the initial permutation-symmetry alone seems to preclude any definition of an initial eigenvalue distribution.

The only assumption underlying our suppression law~\eqref{eq:suppcondgeneral} for general pure states, is the input state's permutation symmetry. This was demonstrated in Sec.~\ref{sec:purestates} for superpositions of Fock product states and entangled states, both obeying the same permutation symmetry and, thus, being subject to the same suppression criterion. Moreover, we highlighted that the input state's permutation symmetry is not necessarily affected by mutual particle distinguishability. Consequently, the suppression of transmission events can persist even in the presence of partially distinguishable particles.

\begin{acknowledgements}
G.W., R.K. and C.D. acknowledge support by the Austrian Science Fund (FWF projects I 2562, P 30459 and M 1849) and the Canadian Institute for Advanced Research (CIFAR, Quantum Information Science Program). C.D. is receiving a DOC fellowship from the Austrian Academy of Sciences. G.D. and A.B. acknowledge support by the EU Collaborative project QuProCS (Grant Agreement No. 641277). Furthermore, G.D. is thankful to the Alexander von Humboldt foundation and M.W. is grateful for financial support from European Union Grant QCUMbER (no. 665148).
\end{acknowledgements}

\begin{appendix}
\section{Proof of inclusion}
\label{appsec:proof1}
Here we show that the extended suppression law for fermions~\eqref{eq:suppcondF} already includes the adapted suppression law~\eqref{eq:Fadaption}. That is, all final particle configurations which are determined to be suppressed according to~\eqref{eq:Fadaption} are also suppressed according to~\eqref{eq:suppcondF}: If the initial particle configuration $\vec{r}$ is invariant under $\mathscr{P}$, either all modes of a given cycle are occupied by one fermion, or none of them are. Hence, for each occupied cycle with length $m_l$, one has to perform $m_l-1$ inversions in order to permute $\vec{r}$ according to $\mathscr{P}$. Furthermore, considering Pauli's exclusion principle \cite{Pauli-ZA-1925}, there are exactly $N_l/m_l$ occupied cycles of length $m_l$. Thus, the number of inversions $w$ in Eq.~\eqref{eq:inversions} is given by $w=\sum_{l=1}^L (m_l-1)N_l/m_l$. That is, according to Eq.~\eqref{eq:Fadaption}, all final particle configurations for which
\begin{align}\label{eq:Appw}
\prod_{\alpha=1}^N \Lambda_\alpha(\vec{s})\neq (-1)^{\sum_{l=1}^L (m_l-1)N_l/m_l}
\end{align}
are suppressed. On the other hand, according to the extended suppression law~\eqref{eq:suppcondF}, the product of all eigenvalues in the final eigenvalue distribution $\Lambda(\vec{s})$ of an allowed state must equal the product of all eigenvalues of the initial eigenvalue distribution $\Lambda_{\mathrm{ini}}$, which yields 
\begin{align*}
\begin{split}
\prod_{\alpha=1}^N \Lambda_\alpha(\vec{s})=&\prod_{l=1}^L\left( \prod_{j=1}^{m_l} e^{i\frac{2\pi j}{m_l}}  \right)^{\frac{N_l}{m_l}}=\prod_{l=1}^L e^{i\frac{\pi N_l(m_l+1)}{m_l}}\\
=&(-1)^{\sum_{l=1}^L (m_l-1)N_l/m_l},
\end{split}
\end{align*}
and reveals that all final configuration $\vec{s}$ which are allowed according to the extended suppression law~\eqref{eq:suppcondF} are also allowed by the adapted law~\eqref{eq:Fadaption}. By contraposition, all configurations $\vec{s}$ which are determined to be suppressed according to~\eqref{eq:Fadaption} are also suppressed according to~\eqref{eq:suppcondF}.

\section{Proof of the extended fermionic suppression law in the wave function-based approach}
\label{appsec:proof2}

To start with, we consider a fermionic initial state~\eqref{eq:state} which is permutation (anti-)symmetric according to Eq.~\eqref{eq:inversions} (that is, the initial particle configuration $\vec{r}$ is invariant under $\mathscr{P}$). It evolves under a unitary transformation matrix as specified in Eq.~\eqref{eq:UA}; 
following Eq.~\eqref{eq:atransform} we obtain
\begin{align*}
\ket{\Psi_\mathrm{evo}^\mathrm{F}(\vec{r})}=\prod_{\alpha=1}^{N}\left(\sum_{k=1}^n U_{d_\alpha(\vec{r}),k} \opd{b}_k\right)\ket{0}.
\end{align*}
The overlap with a final state $\ket{\Psi^\mathrm{F}(\vec{s})}$ (defined in Eq.~\eqref{eq:statefinal}) can then be expressed as a sum over elements $\sigma$ of the symmetric group $S_N$: 
\begin{align}\label{eq:summands}
\bracket{\Psi^\mathrm{F}(\vec{s})}{\Psi_\mathrm{evo}^\mathrm{F}(\vec{r})}
=&\sum_{\sigma\in\mathrm{S}_N} T_\sigma,\\
T_\sigma=&\ \sgn(\sigma) \prod_{\alpha=1}^N U_{d_\alpha(\vec{r}),  d_{\sigma(\alpha)}(\vec{s})}.\notag
\end{align}
We now make the following hypothesis (H): \textit{For all $\sigma\in\mathrm{S}_N$, one can find a pair of distinct particles
$\mu$ and $\nu$ initially occupying modes $d_\mu(\vec{r})$ and $d_\nu(\vec{r})$ which belong to the same cycle of $\pi$ and such that $\lambda_{d_{\sigma(\mu)}(\vec{s})}=\lambda_{d_{\sigma(\nu)}(\vec{s})}$. }

Letting $m_l$ be the length of the cycle, there exists a $\kappa\in\{1,\dots,m_l-1\}$ such that 
\begin{align}
\label{eq:munu}\pi^\kappa(d_\mu(\vec{r}))&=d_\nu(\vec{r}).
\end{align}
Utilizing the permutation phase relation~\eqref{eq:sym} $\kappa$ times, together with~\eqref{eq:munu}, yields 
\begin{align}\label{eq:Uelementsym}
U_{d_\nu(\vec{r}),k}=&U_{d_\mu(\vec{r}),k}\ \lambda_k^\kappa \exp\left(i[\theta(d_\nu(\vec{r}))-\theta(d_\mu(\vec{r}))]\right).
\end{align}
We now consider the permutation $\sigma'$ obtained by composing  $\sigma$  with the transposition of $\mu$ and $\nu$, that is $\sigma'(\mu)=\sigma(\nu)$, $\sigma'(\nu)=\sigma(\mu)$ and $\sigma'(\alpha)=\sigma(\alpha)$ for $\alpha\neq\mu,\nu$.  
Using Eq.~\eqref{eq:Uelementsym} and $\sgn(\sigma')=-\sgn(\sigma)$, we find for the summands of Eq.~\eqref{eq:summands}
\begin{align*}
T_\sigma=-\left(\frac{\lambda_{d_{\sigma(\mu)}(\vec{s})}}{\lambda_{d_{\sigma(\nu)}(\vec{s})}}\right)^\kappa T_{\sigma'},
\end{align*} 
and by our hypothesis, $T_\sigma=-T_{\sigma'}$. The summands in Eq.~\eqref{eq:summands} therefore  cancel two by two and 
 $\bracket{\Psi^\mathrm{F}(\vec{s})}{\Psi_\mathrm{evo}^\mathrm{F}(\vec{r})}=0$, i.e.
\begin{align*}
  \mathrm{(H)}\quad \Rightarrow \quad \vec{s} \ \text{ is suppressed.}
\end{align*} 
 
By contraposition, if $\vec{s}$ is allowed, then there exists a permutation $\sigma\in\mathrm{S}_N$ such that for all cycles of $\pi$ and for all pairs of distinct particles $(\mu,\nu)$ initially in modes belonging to the same cycle, $\lambda_{d_{\sigma(\mu)}(\vec{s})}\neq\lambda_{d_{\sigma(\nu)}(\vec{s})}$.
It follows that each initially populated cycle of length $m_l$ contributes $m_l$ distinct eigenvalues to the final eigenvalue distribution. However, recalling the  discussion above Eq.~\eqref{eq:Ndef}, these  eigenvalues must also be $m_l$th roots of unity. Therefore, each initially populated cycle of $\pi$ with length $m_l$ contributes all the distinct $m_l$th roots of unity to the final eigenvalue distribution, and by the definition of the initial eigenvalue distribution $\Lambda_\mathrm{ini}$ above Eq.~\eqref{eq:suppcondF} we conclude that
\begin{align*}
\vec{s}\ \text{ is allowed} \quad \Rightarrow \quad \Lambda(\vec{s})= \Lambda_\mathrm{ini}.
\end{align*}
The extended suppression law~\eqref{eq:suppcondF} follows by contraposition.

\end{appendix}

%----------------------------------------------------------------------------------
%                Bibliography
%----------------------------------------------------------------------------------
\bibliographystyle{aipnum4-1} % Tell bibtex which bibliography style to use
%\bibliography{CDbibSymSupp}


\end{document}